\documentclass[useAMS,usenatbib]{mnras}

\usepackage{hyperref}
\usepackage{graphicx}

\title[Modelling the CSED]{Modelling the cosmic spectral energy distribution and extragalactic background light over all time}
\author[Andrews et al.]{S. K. Andrews$^{1, 2}$\thanks{E-mail: stephen.andrews@icrar.org}, S. P. Driver$^{1, 2}$, L. J. M. Davies$^{1}$, C. d. P. Lagos$^{1}$, A. S. G. Robotham$^{1}$ \\
$^{1}$International Center for Radio Astronomy Research, The University of Western Australia, 35 Stirling Highway, Crawley, WA 6009, Australia \\
$^{2}$School of Physics \& Astronomy, University of St Andrews, North Haugh, St Andrews, KY16 9SS, United Kingdom }
\begin{document}

\date{Accepted 1988 December 15. Received 1988 December 14; in original form 1988 October 11}

\pagerange{\pageref{firstpage}--\pageref{lastpage}} \pubyear{2002}

\maketitle

\label{firstpage}

\begin{abstract}
We present a phenomological model of the Cosmic Spectral Energy Distribution (CSED) and the integrated galactic light (IGL) over all cosmic time. This model, based on an earlier model by Driver et al. (2013), attributes the cosmic star formation history to two processes -- firstly, chaotic clump accretion and major mergers, resulting in the early-time formation of bulges and secondly, cold gas accretion, resulting in late-time disc formation. Under the assumption of a Universal Chabrier initial mass function, we combine the Bruzual \& Charlot (2003) stellar libraries, the Charlot \& Fall (2000) dust attenuation prescription and template spectra for emission by dust and active galactic nuclei to predict the CSED -- pre- and post-dust attenuation -- and the IGL throughout cosmic time. The phenomological model, as constructed, adopts a number of basic axioms and empirical results and has minimal free parameters. We compare the model output, as well as predictions from the semi-analytic model \textsc{galform} to recent estimates of the CSED out to $z=1$. By construction, our empirical model reproduces the full energy output of the Universe from the ultraviolet to the far-infrared extremely well. We use the model to derive predictions of the stellar and dust mass densities, again finding good agreement. We find that \textsc{galform} predicts the CSED for $z < 0.3$ in good agreement with the observations. This agreement becomes increasingly poor towards $z = 1$, when the model CSED is $\sim$50 per cent fainter. The latter is consistent with the model underpredicting the cosmic star formation history. As a consequence, \textsc{galform} predicts a $\sim$30 per cent fainter IGL.
\end{abstract}


\section{Introduction}
\label{sec:intro}

One of the goals of modern cosmology is to understand, explain and predict the evolution of mass, energy, and structure from the Big Bang to the present day. Here we focus on the evolution of energy emanating from the ultraviolet to far-infrared wavelength range -- photon production, transport, and redistribution -- over a 13 billion year timeline (i.e., since recombination). The energy content of the Universe over this wavelength range is often expressed through three measurable quantities: the cosmic spectral energy distribution (CSED) as a function of redshift, the extragalactic background light (EBL), and the integrated galactic light (IGL). These quantities are each measurable via distinct direct or indirect methods, but are also closely related. The CSED (e.g. \citealt{driver08,dominguez11,driver16a,andrews16b}) describes the photons generated within a cosmologically representative volume at a specific epoch. The EBL can therefore be expressed as the volume and luminosity distance weighted integral of the redshifted CSED since decoupling. Hence, the EBL describes the distribution of photons observed today, originating from all post-decoupling processes from $\lambda_\mathrm{obs} = 100$~nm to $\lambda_\mathrm{obs} = 1$~mm. The IGL represents the dominant component of the EBL and is best derived from number counts of discrete resolved objects (e.g. galaxies and AGN) as observed by a sufficiently deep extragalactic survey (e.g. \citealt{driver16b}). In the absence of any diffuse light, the EBL and IGL will be identical -- hence, any discrepancy between the two constrains the level of diffuse emission \citep{driver16b}.

\newpage

The CSED, EBL and IGL encode statistical information about past and ongoing processes of photon production. The primary source of new photons at low and intermediate redshifts is from stellar nucleosynthesis. However, approximately half the photons being produced by stars are absorbed by dust in the host galaxy, which heats and re-radiates in the far infrared \citep{savage79,driver08,driver16a,andrews16b}. The remaining photons escape into the intergalactic medium. At higher redshifts, both components (direct and reprocessed photons) are supplemented by mass accretion onto active galactic nuclei (AGN) and their surrounding dust torii (e.g. \citealt{richards06}). Again, approximately half of the photons produced by AGN are also attenuated. However, in the AGN case the dust torus is significantly hotter and hence re-radiates at mid-infrared wavelengths. Remarkably, this approximate equality between optical and mid- to far-infrared energy output persists when integrating over all redshifts and is often divided into two distinct components -- the cosmic infrared background (8~$\mu$m $< \lambda_\mathrm{obs} < 1$~mm in the observed frame; \citealt{dwek98}); and the cosmic optical background ($0.1~\mu$m$~< \lambda_\mathrm{obs} < 8~\mu$m), see e.g. \citet{driver16b}. 

Given the intimate link between the CSED, the EBL and galaxy evolution, it is insightful to obtain CSED and EBL predictions from semi-analytic (e.g. \citealt{gilmore12,somerville12,inoue13}) and phenomological models (e.g. \citealt{partridge67a,partridge67b,finke10,driver13,khaire15}) of galaxy formation. Semi-analytic models, e.g. \citet{lacey16}, begin with dark matter haloes as output from some $n$-body simulation (e.g. \citealt{springel05}), which accrete gas from the intergalactic medium according to a predefined prescription. The model determines the star formation history, metallicity evolution and feedback from supernovae and active galactic nuclei from the accretion rate and the halo merger tree. The unattenuated SED can then be computed using stellar population synthesis codes (e.g. \citealt{bc03,maraston05}), with dust attenuation and re-emission added via prescription.

Phenomological models are empirically driven and can be divided into two classes: forward modelling and backward modelling. Forward models (e.g. \citealt{driver13}) are based on measurements or assumptions of:

\begin{itemize}
\item The cosmic star formation history (CSFH; e.g. \citealt{hopkins06,madau14});
\item Stellar population synthesis (e.g. \citealt{bc03,maraston05})
\item A known Universal initial mass function (IMF);
\item The unattenuated spectral energy distribution (SED) of starlight; and 
\item A dust attenuation and emission model. 
\end{itemize}

Backward modelling arises from evolving observed properties of nearby galaxies, for example luminosity functions \citep{franceschini08} or the classification of galaxies by SED fitting \citep{dominguez11}, and propagating them backwards in time. Due to their simplistic design and high-level approach, phenomological models have fewer free parameters than approaches which look to encode galaxy physics and also recover clustering signatures -- i.e. flexible but limited (phenomological) versus comprehensive but immutable (semi-analytic, hydrodynamic and $n$-body simulations).

Direct measurements of the cosmic optical (e.g. \citealt{matsumoto05,matsumoto15,levenson08,matsuoka11,hess13,biteau15,magic16,zemcov17,matsuura17,mattila17}) and infrared (e.g. \citealt{puget96,fixsen98,hauser98}) backgrounds have been made using a number of different facilities and methods. Likewise, the IGL has been measured in the optical (e.g. \citealt{madau00,totani01,xu05,keenan10}) to the far-infrared (e.g. \citealt{bethermin12,carniani15}, with some groups measuring how the cosmic infrared background accumulates as a function of redshift at specific wavelengths (e.g. \citealt{chary01,marsden09,jauzac11,berta11,bethermin12b,viero13}). These measurements are usually confined to a single facility and hence a narrow portion of the spectrum.

Robust measurements of the broader CSED have only recently become possible, as they rely on the availability of sufficiently deep and wide multiwavelength data. The first measurement of the dust corrected CSED and hence the far-infrared energy output was made by \citet{driver08} from existing optical luminosity density measurements. This measurement was updated by \citet{driver12}, who integrated luminosity functions from FUV through $K$ derived from the Galaxy and Mass Assembly (GAMA) survey. \citet{dominguez11} used a SED fitting code in order to interpolate between broadband filters, \citet{kelvin14} examined the CSED as a function of galaxy morphology and \citet{driver16a} extended observations to the far-infrared. Finally, \citet{andrews16b} measured the CSED out to $z=1$ based on a completely homogenous data reduction procedure across the entire ultraviolet to far-infrared wavelength range. In a similar vein, \citet{driver16b} computed the extrapolated IGL over the same wavelength range using consistent analysis techniques. 

With this data now in hand, this work aims to both extend the \citet{driver13} forward model as well as derive a prediction of the CSED and IGL from the \textsc{galform} semi-analytic model and compare each to the observations. In Section \ref{sec:empirical}, we briefly recap our measurements of the IGL and CSED as reported by \citet{driver16b} and \citet{andrews16b} respectively. In Section \ref{sec:toymodel} we revisit the \citet{driver13} model and extend it to include both dust attenuation and AGN. In Section \ref{sec:csed}, we derive predictions for both the unattenuated and attenuated CSEDs. In Section \ref{sec:sams}, we obtain predictions of the CSED from two recent incarnations of \textsc{galform}. In Section \ref{sec:results}, we compute a number of useful relationships from our phenomological model, including the integrated photon escape fraction, the comoving IGL, and the buildup of dust and stellar mass densities as predicted by our model. We conclude in Section \ref{sec:conclusion}.

We use AB magnitudes and assume $H_0 = 70~h_{70}~\mathrm{km~s^{-1}~Mpc^{-1}}$, $\Omega_\Lambda = 0.7$ and $\Omega_m = 0.3$ throughout.

\section{Empirical estimates of the IGL and CSED}
\label{sec:empirical}

The genesis of the IGL and CSED data we wish to reproduce is critical to our analysis, so we briefly describe how these data were derived below.

\subsection{The integrated galactic light (IGL)}

In \citet{driver16b}, we measured the IGL from a compendium of deep and wide galaxy number count data from the far-ultraviolet to the far-infrared. This included the GAMA \citep{wright16a} and G10/COSMOS \citep{andrews16a} datasets as well as a selection of deeper surveys, including data from the \textit{Galaxy Evolution Explorer}, \textit{Hubble}, \textit{Spitzer}, \textit{Wide-field Infrared Survey Explorer} and \textit{Herschel} space telescopes. We found that the contributions to the IGL as a function of magnitude are bound at both bright and faint magnitudes for all bands. To determine the total energy in each band, we fitted a spline to the luminosity-weighted number count data and integrated it. Finally, to obtain a physically motivated IGL spectrum, we used a preliminary version of the model described in this work. Summing our IGL spectrum over the relevant wavelength ranges, we found values for the cosmic optical and infrared backgrounds of $24 \pm 4$ and $26 \pm 5$~nW~m$^{-2}$~sr$^{-1}$ respectively. These extrapolated IGL measurements agree well with previous literature IGL measurements, but fall significantly below the direct EBL estimates based on absolute background analysis of \textit{Hubble} data (e.g. \citealt{bernstein02}). However, our IGL measurements, with a 20 per cent adjustment upward to account for intra-halo or intra-cluster light and low surface brightness galaxies, agree closely with the high-energy gamma ray constraints from H.E.S.S. \citep{hess13} and MAGIC \citep{magic16}. Our measurements also agree with the direct EBL observations by the \textit{Pioneer} and \textit{New Horizons} spacecraft taken from the outer solar system, where the impact of zodiacal light is much less significant \citep{matsuoka11,zemcov17}. This suggests that the direct measurements of the cosmic optical background may have underestimated the terrestrial (Earthshine) and/or zodiacal contributions to the total sky background (see e.g. \citealt{kawara17}). In summary, the IGL measurements presented in \citet{driver16b} appear robust at all wavelengths with a tentative detection of diffuse light at the 0--20 per cent level in the near-infared.

\subsection{The CSED as a function of redshift}

While the EBL describes the total integrated light along the path length of the Universe, the (total) CSED represents its subdivision by redshift. In \citet{andrews16b}, we measured the (resolved) CSED from the far-ultraviolet to the far-infrared over a 7.5 Gyr baseline using multiwavelength imagery and spectroscopy from the GAMA \citep{driver11} and COSMOS \citep{scoville07} surveys. The GAMA spectroscopic and imaging data are described in \citet{liske15} and \citet{driver16a} with photometry performed by \citet{wright16a}, while the assembly of redshift and photometric data in the COSMOS region are described by \citet{davies15} and \citet{andrews16a} respectively. 

The photometric measurements were then interpolated using the SED fitting code \textsc{magphys} (Multi-wavelength Analysis of Galaxy Physical Properties; \citealt{dacunha08}) by \citet{driver17}. As a reminder, the \textsc{magphys} fits use the \citet{bc03} stellar libraries based on a \citet{chabrier03} IMF and the two component \citet{charlot00} dust attenuation model. The resulting far-infrared dust emission spectrum comprises of polycyclic aromatic hydrocarbons, warm dust and two populations of cold dust in thermal equlibrium with varying temperatures. The resulting SED fits are generally robust, producing stellar mass and star formation rate estimates in line with independant estimates (\citealt{taylor11,davies16}, see also the direct comparison in \citealt{driver17}).

We stacked the \textsc{magphys} fits for individual galaxies in ten redshift bins to derive both the unattenuated (dust corrected) and attenuated (observed) CSED for $0.02 < z < 0.99$. We found that the total energy output declined from $(5.1 \pm 1.0) \times 10^{35} h_{70}$~W~Mpc$^{-3}$ at $z=0.905$ to $(1.3 \pm 0.3) \times 10^{35} h_{70}$~W~Mpc$^{-3}$ at the current epoch, a rate slightly slower than the well known decline in cosmic star formation \citep{lilly96} -- as one would expect given that the CSED samples both young and old starlight. The quoted uncertainty in the absolute normalisation of the CSEDs ranges from 11 to 27 per cent, and is dominated by cosmic sample variance. Additional, semi-quantifiable and wavelength-dependent uncertainties arise from SED modelling error, incompleteness and the lack of complete far-infrared and ultraviolet measurements in G10/COSMOS. These are discussed in \citet{andrews16b}, their Figure 6 and Section 3.3.

\section{Reconstructing the CSED}
\label{sec:toymodel}

We now look to build a phenomological model of the CSED and elect to follow the pathway laid out in \citet{driver13} with the addition of active galactic nuclei (AGN) and a more comprehensive dust modelling prescription. To build our revised phenomological model of the energy output of the Universe, one needs:

\begin{enumerate}
\item Empirical estimates of the cosmic star formation history (CSFH; e.g. \citealt{hopkins06,madau14,driver17})
\item The assumption of a mean initial mass function (IMF) across cosmic time (e.g. \citealt{kennicutt83,kroupa01,baldry03,chabrier03})
\item A stellar population model (e.g. \citealt{bc03,leborgne04,maraston05})
\item A description of how the metallicity of newly formed stars evolves with redshift (\citealt{tremonti04,erb06,zahid11} following \citealt{driver13})
\item A dust attenuation model (e.g. \citealt{cardelli89,calzetti00,charlot00,gordon03}) and corresponding emission spectra in the far infrared (e.g. \citealt{chary01,zubko04,draine07,dale14})
\item An emission model from active galactic nuclei (e.g. \citealt{elvis94,dale14,siebenmorgen15}) and either a prescription of linking that to the spheroid star formation rate, or some description of how AGN activity evolves with time (e.g. \citealt{richards06,palanque16}).
\end{enumerate}

Constructing a model in this manner should, in theory, require no tunable initial conditions -- the choice of parameters in stellar, dust and AGN modelling should all be guided by empirical data. However, free parameters are unavoidable when the available data is insufficient to constrain the model. At the heart of the \citet{driver13} two-phase model there are two axioms, which we also adopt here: 

\begin{enumerate}
\item AGN activity traces stellar mass growth in spheroids
\item the formation of material ending up in spheroids today dominates at high redshift
\end{enumerate}

The first axiom is based on the $M_{bh}-M_{*, sph}$ and $M_{bh}-\sigma$ relations \citep{magorrian98,gebhardt00}, which implies co-evolution between the central black hole and the surrounding spheroid \citep{silk98,kormendy13,graham15}. However, while the $M_{bh}-M_{*, sph}$ relation is irrefutable, multiple studies have found no or only weak correlation between total (i.e. disc and spheroid) star formation and AGN activity \citep{rosario12,mullaney12a,stanley15}. At $z \la 1.6$, we expect this to be the case as total star formation is dominated by cold gas accretion that never reaches the latent supermassive black hole (i.e. it is confined to discs only). This redshift corresponds to the transition point in the \citet{driver13} model. At $z \ga 1.6$, we explicitly link \textit{spheroid} star formation with AGN activity -- both are associated with scenarios in which low angular momentum gas is driven towards the centre of the galaxy, fuelling a central starburst and black hole accretion. The lack of correlation at high redshift is therefore surprising. One possible solution is that these studies are either failing to separate bulge and disc star formation or mismatching timescales. Instantaneous measurements may be affected by an AGN lag and/or the stochasticity of short-term star formation. When considering time-averaged AGN activity (as determined by X-ray stacking) and far-infared star formation measurements the correlation appears to strengthen \citep{mullaney12b,chen13,rodighiero15}. 

The case for the second axiom is stronger. The recent advent of integral field spectroscopic observations has enabled more robust determination of the mass growth history of galaxies. \citet{mcdermid15} found that early type galaxies have formed 50 per cent of their stars by $z = 3$ and 90 per cent by $z \sim 0.9$. A similar picture is painted by \citet{ibarramedel16}, with red, quiescent, high mass and/or early type galaxies forming a greater portion of their stars at higher redshifts. In both studies, low-mass galaxies undergo steady mass growth to the current epoch. \citet{gonzalez15} showed that early-type galaxies contain older stellar populations ($\log(\mathrm{age}) \sim 9.5-10$) than late type galaxies ($\log(\mathrm{age}) \sim 8.5-9.5$). Sa and Sb types show an older stellar population at low galactocentric radii, potentially corresponding to bulge material. Early type and high-mass galaxies contain the majority of spheroid material in the Universe, while low-mass galaxies consist of predominantly disc material (e.g. \citealt{moffett16}). This is consistent with the scenario we envisage in the two-phase model -- spheroid star formation occuring primarily at higher redshifts, while star formation linked to cold gas accretion continues in low-mass galaxies and discs to the current epoch.

We proceed under the assumption that AGN activity is causally linked to, or coincidental with spheroid star formation on a time-averaged basis and now build up our model over five key stages:

\begin{enumerate}
\item Compute the unattenuated CSED from the CSFH and an adopted gas-phase metallicity evolution.
\item Compute the attenuated CSED from a set of selected dust attenuation parameters, given the CSFH and metallicity evolution in stage 1.
\item Add far-infrared spectra based on the attenuated energy derived in stage 2.
\item Add AGN spectra based on the evolution of the AGN luminosity function with redshift.
\item Construct the IGL from the full CSED as a function of redshift.
\end{enumerate}

These ingredients are described below.

\subsection{The cosmic star formation history}

\begin{figure*}
\begin{minipage}{6.5in}
\begin{center}
\includegraphics[width=0.99\linewidth]{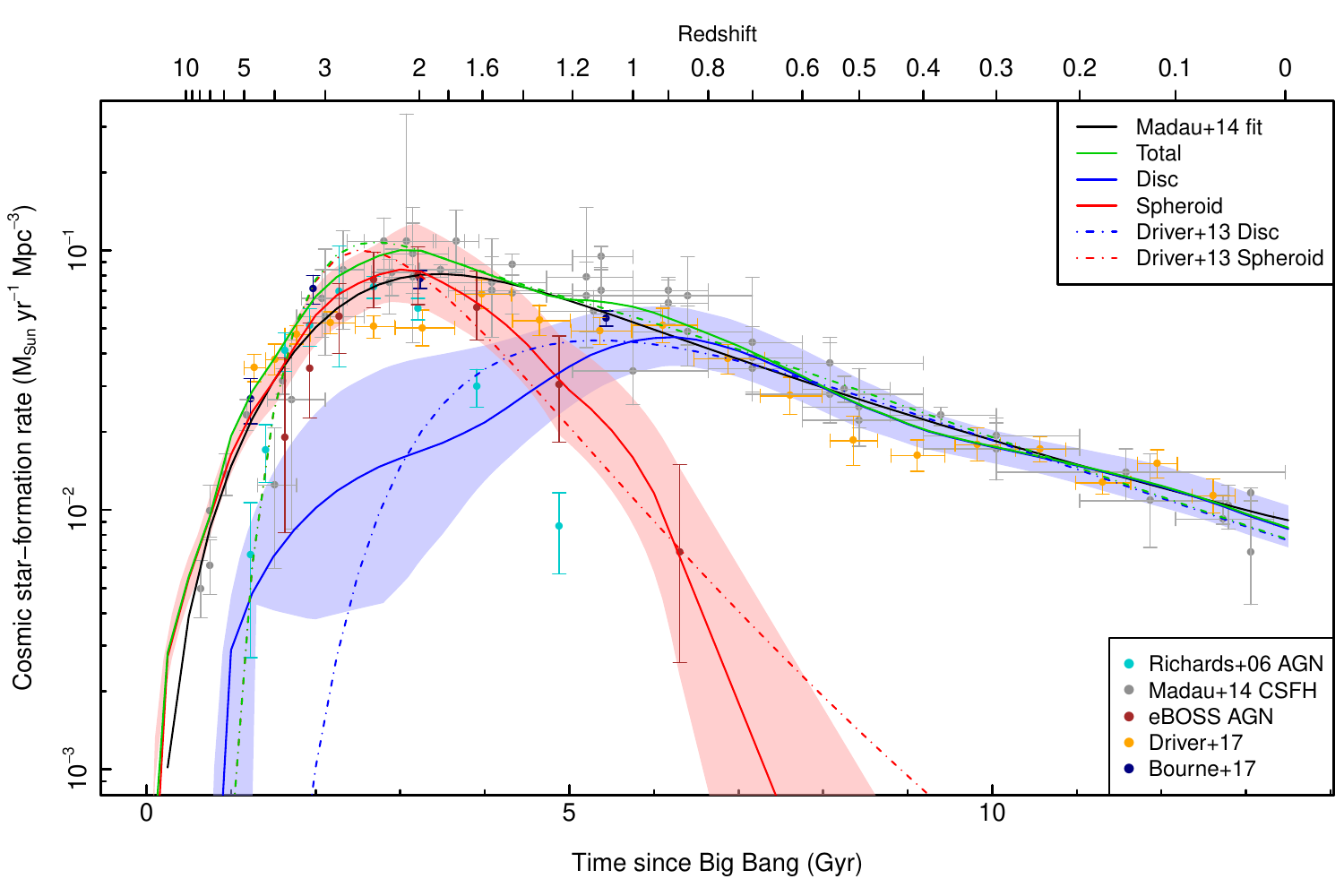}
\caption{Our fitting functions to the CSFH for discs (blue), spheroids (red) and combined (green) compared to those for \citet{driver13} (dashed line) and \citet{madau14} (solid black line). The corresponding error regions in the disc and spheroid CSFHs are also shown. The \citet{madau14} CSFH data (grey points), the \citet{driver17} and \citet{bourne17} CSFH data (orange and dark blue points respectively) and the \citet{richards06} AGN luminosity function data scaled to the \citet{madau14} fitting function (red points) are also shown.}
\label{fig:csfh}
\end{center}
\end{minipage}
\end{figure*}

We take the opportunity to update the \citet{driver13} fitting functions of the CSFH as follows:

We replace the compilation of cosmic star formation rate measurements with that from \citet{madau14} augmented by the recent measurements from \citet{driver17} and \citet{bourne17}. We replace the \citet{richards06} AGN $i$ band total luminosity data with $g$ band (rest frame) data from the extended Baryon Oscillation Spectroscopic Survey (eBOSS; \citealt{palanque16}). We calculated the total quasar luminosity and error from the 16th, 50th and 84th percentile of 1000 Monte Carlo integrations of the pure luminosity-evolution model of the \citet{palanque16} quasar luminosity function.

We scale the peak of the eBOSS total quasar luminosity, which occurs at $z = 2.01$, to the \citet{madau14} fitting function. From this, we determine the CSFH for spheroids by fitting an eight-point spline weighted by the inverse relative error squared using the compiled CSFH data for $z > 2.01$ and the eBOSS AGN data below this redshift (red line on Figure \ref{fig:csfh}). We elect to use the compiled CSFH data above the peak redshift because it is more comprehensive and precise than the AGN data -- in the adopted model, about half of spheroid stellar mass forms prior to $z = 2.01$.

The eBOSS data represent optical luminosities, and thus do not capture activity from obscured AGN. Under the assumptions of the model, an obscured fraction which evolves with redshift changes how quickly spheroid star formation activity declines since $z \sim 2$ if we maintain that the spheroid CSFH is equal to the total CSFH prior to that redshift. The evolution of the obscured AGN fraction as a function of redshift is still unclear. \citet{triester06} find from quasar x-ray luminosity functions that the obscured fraction evolves as $(1+z)^{0.3-0.4}$, while \citet{lusso13}, also using x-ray data, find no evidence for evolution. The shape of the spheroid CSFH has direct consequences for the bulge to disc stellar mass ratio, and corresponding CSEDs, at $z = 0$. Adopting the evolution in \citet{triester06} will result in a steeper drop-off in the spheroid CSFH, less spheroid stellar mass and a fainter spheroid CSED at $z = 0$. Conversely, adopting an x-ray luminosity density evolution that declines more slowly, as in, e.g. \citet{ueda14} results in the opposite. Similarly, adopting an earlier (later) peak will result in less (more) spheroid stellar mass and redder, fainter (bluer, brighter) emission at $z = 0$. Accurate bulge-disc decomposition over the $0.1 < z < 2$ range, and corresponding stellar mass estimates, will test these scenarios and allow the reconstruction of the disc and spheroid CSFHs without the need to adopt AGN activity as a proxy.

The CSFH for discs (solid blue curve in Figure \ref{fig:csfh}) is determined by fitting an eight point spline to the \citet{madau14} CSFH data minus the CSFH for spheroids. Each spline is then extrapolated to cover lookback times between 0 and 13.5 Gyr. We also impose a CSFH floor of $10^{-4} M_\odot$~yr$^{-1}$~Mpc$^{-3}$ to reduce the numerical integration error arising when deriving the stellar mass density in spheroids. 

The resulting CSFH functions and data are shown in Figure \ref{fig:csfh}. The transition from spheroid star formation to disc star formation happens $\sim$1~Gyr later ($z \sim 1.2$ compared to $z \sim 1.7$) in this model compared to \citet{driver13}. This is due to the use of the eBOSS AGN data, and a higher CSFH for spheroids at very high redshift because we fit to the CSFH data instead of the AGN data. Combined with using spline fitting to avoid assuming a functional form, this improves the fitting of the CSFH at high redshifts over the \citet{driver13} model. The precise shape of the disc CSFH at $t \sim 5$~Gyr is, to some degree, a byproduct of the spline fitting and subtraction and has little physical significance. Otherwise, the fitting functions are similar to those adopted by \citet{driver13} for the CSFH for spheroids and discs (dashed lines).

We estimate the error associated with the CSFH by repeating the spline fitting procedure for the lower and upper bounds of the individual CSFH data points and show the error regions on Figure \ref{fig:csfh}. The resulting $\approx$40 per cent error in the total CSFH at $z = 0$ represents a conservative estimate of the error in the CSFH by itself. It is more reasonable as an estimate of total model error, which incorporates uncertainties in the model assumptions, the initial mass function, stellar libraries and gas phase metallicity.

\subsection{Model stellar populations}

\begin{figure}
\begin{center}
\includegraphics[width=0.99\linewidth]{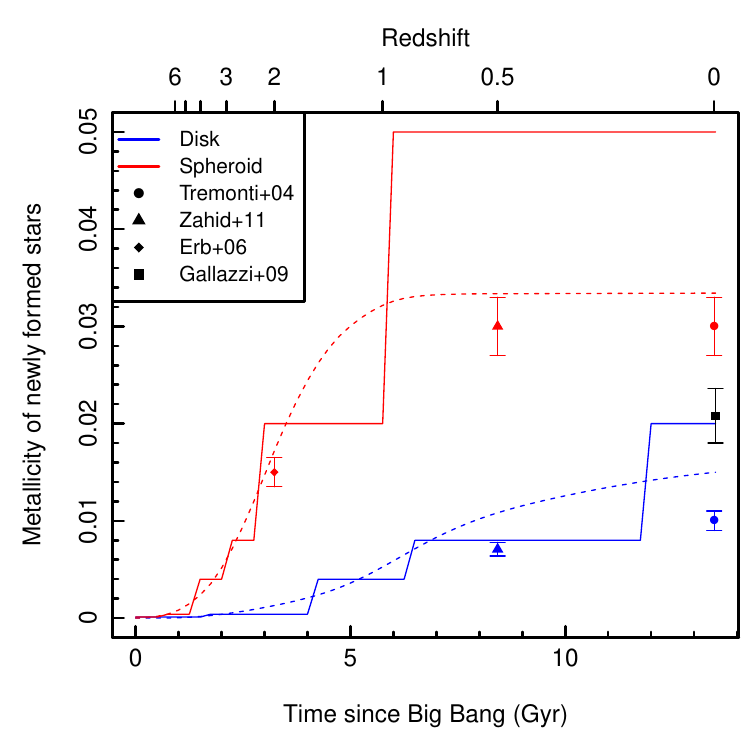}
\caption{The adopted metallicity of newly formed stars for spheroids (red) and discs (blue). Metallicity curves in the absence of rounding are shown in dashed lines. The inferred metallicities of spheroids and discs, from \citet{driver13} based on the underlying \citet{tremonti04,erb06} and \citet{zahid11} data, as well as the cosmic stellar phase metallicity \citep{gallazzi09} are also shown.}
\label{fig:metallicity}
\end{center}
\end{figure}

\begin{figure}
\begin{center}
\includegraphics[width=0.99\linewidth]{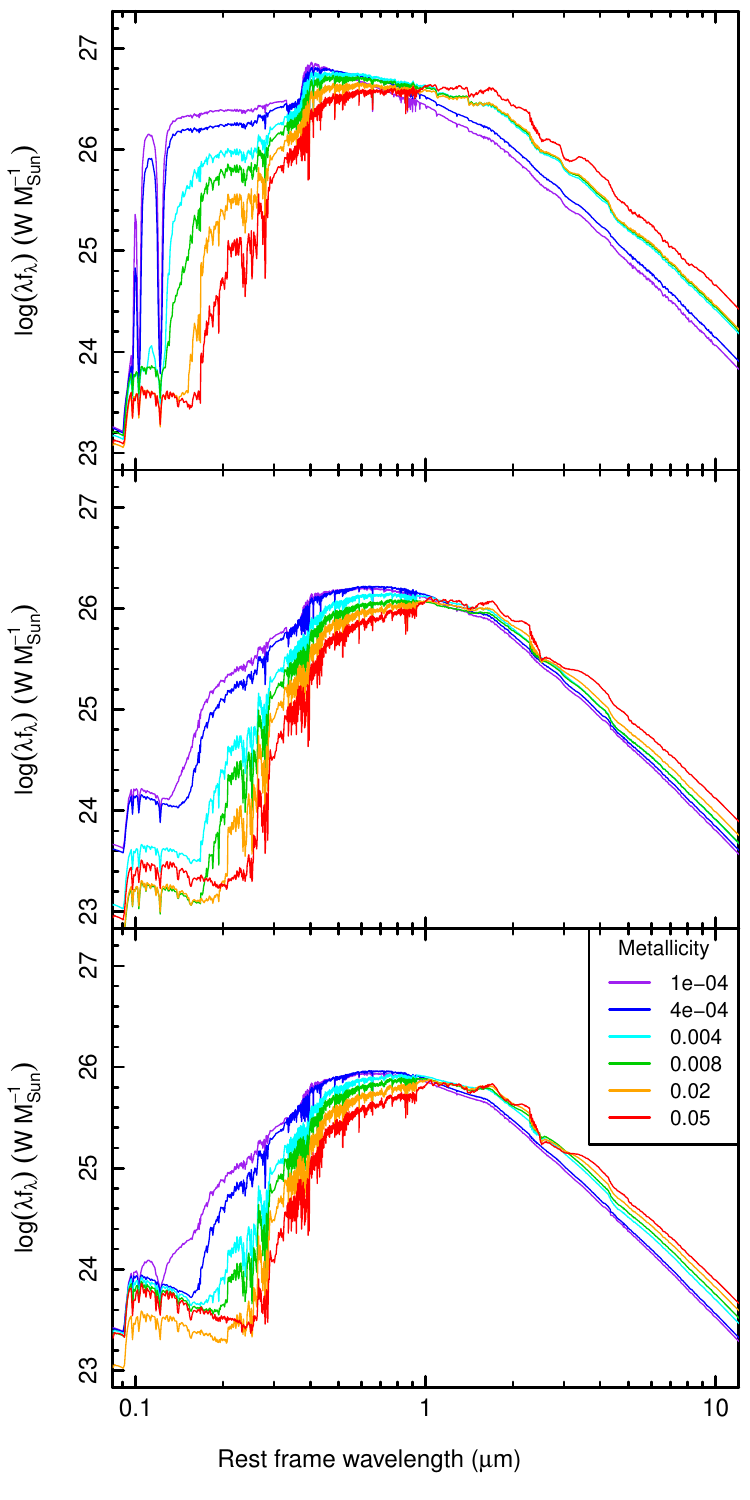}
\caption{Unattenuated SEDs for simple stellar populations constructed using the CSFH for discs 8~Gyr after the Big Bang with the indicated metallicities at an age of 1~Gyr (top), 5~Gyr (centre) and 10~Gyr (bottom). The units on the Y axis are watts per solar mass of stars formed.}
\label{fig:metallicity_spectra}
\end{center}
\end{figure}

To model the stellar population, we use the \textsc{galaxev} software and the \citet{bc03} stellar models for consistency with the \citet{andrews16b} empirical CSED measurement. Both the \citet{bc03} and \textsc{PEGASE} 2 \citep{leborgne04} model used by \citet{driver13} employ the same set of Padova (1994) isochrones, but differ in their underlying library of stellar atmospheres -- \citet{bc03} uses the theoretical BaSeL library \citep{allard95} while \textsc{PEGASE} 2 uses the empirical ELODIE library \citep{prugniel01}. The difference between the two stellar libraries should be small over the relevant wavelength range \citep{conroy10}. A detailed exploration of the effect of assuming different stellar libraries on the empirical and model CSEDs would require refitting SEDs to subsamples of the \citet{driver17} catalogue and is outside the scope of this paper. Regardless, any differences are not likely to be significant in light of measurement and other model uncertainties. If the BC03 libraries are found to be insufficiently accurate, the CSED measurements and models will need to be revised accordingly.

We use a \citet{chabrier03} IMF, again to be consistent with the empirical CSED measurement. (The \citet{driver13} model uses a \citet{baldry03} IMF, which can be converted to a Chabrier IMF by scaling the CSFH appropriately.)

We compute model spectra in 0.25~Gyr timesteps using the adopted CSFH splines evaluated at 10~Myr intervals. We adopt a metallicity of newly formed stars that increases linearly with stellar mass formed for each component with no time lag (i.e. instantaneous enrichment), similar to \citet{driver13}. However, unlike \textsc{PEGASE} 2 which has a broad range of metallicities, the \citet{bc03} models are only computed for six different discrete metallicities -- $Z = 0.0001$, 0.0004, 0.004, 0.008, 0.02 ($Z_\odot$) and 0.05. We do not interpolate beween metallicities which is considered bad practice, but round to the nearest available \citet{bc03} model (see Figure \ref{fig:metallicity}). We also require the metallicity to reach $Z_\odot$ at $t=3.25$~Gyr for spheroids and 0.75~$Z_\odot$ at $t = 13.5$~Gyr for discs. This choice is guided by the \citet{driver13} model, which reaches metallicities of 0.01 and 0.03 for spheroids and discs respectively and the underlying \citet{tremonti04,erb06} and \citet{zahid11} data, but sets a higher metallicity for discs. The \citet{driver13} prescription reaches a cosmic stellar phase metallicity significantly lower than the $1.04 \pm 0.14 Z_\odot$ at the current epoch \citep{gallazzi09}, so we adopt a higher trajectory for the disc metallicity. The adopted metallicity evolution, with and without rounding, is shown in Figure \ref{fig:metallicity}.

For illustration purposes, Figure \ref{fig:metallicity_spectra} shows SEDs for simple stellar populations with ages 1, 5 and 10~Gyr as a function of metallicity. The input CSFH is that for discs at 8~Gyr after the Big Bang. While the model CSEDs vary considerably in the ultraviolet and short-wavelength optical, the total flux output varies by only 0.3~dex in the near-infrared between the most extreme metallicities. The least scatter occurs in the rest-frame $Y$ band -- at $1~\mu$m, flux ranges from $10^{26.44}$~W per solar mass of stars formed ($Z = 0.0001$) to $10^{26.64}$~W~$M_\odot^{-1}$ ($Z = 0.004$) at $t = 1$~Gyr. The spread decreases to between $10^{26.07}$~W~$M_\odot^{-1}$ ($Z = 0.0004$) to $10^{26.14}$~W~$M_\odot^{-1}$ ($Z = 0.008$) at $t = 5~$Gyr and $10^{25.82}$~W~$M_\odot^{-1}$ ($Z = 0.004$) to $10^{25.91}$~W~$M_\odot^{-1}$ ($Z = 0.05$) at $t = 10$~Gyr. 

From our adopted CSFH, IMF and metallicity evolution, 61 and 74 per cent of stellar mass forms at metallicities between 0.004 and 0.02 for both spheroids and discs, where the near-infrared flux difference is negligible. Imprecision in modelling metallicities may affect the blue portion of the spheroid CSED notably, but the impact on the overall CSED is minimal. 

\subsection{Dust modelling}
\label{sec:dust}

\begin{figure}
\begin{center}
\includegraphics[width=0.99\linewidth]{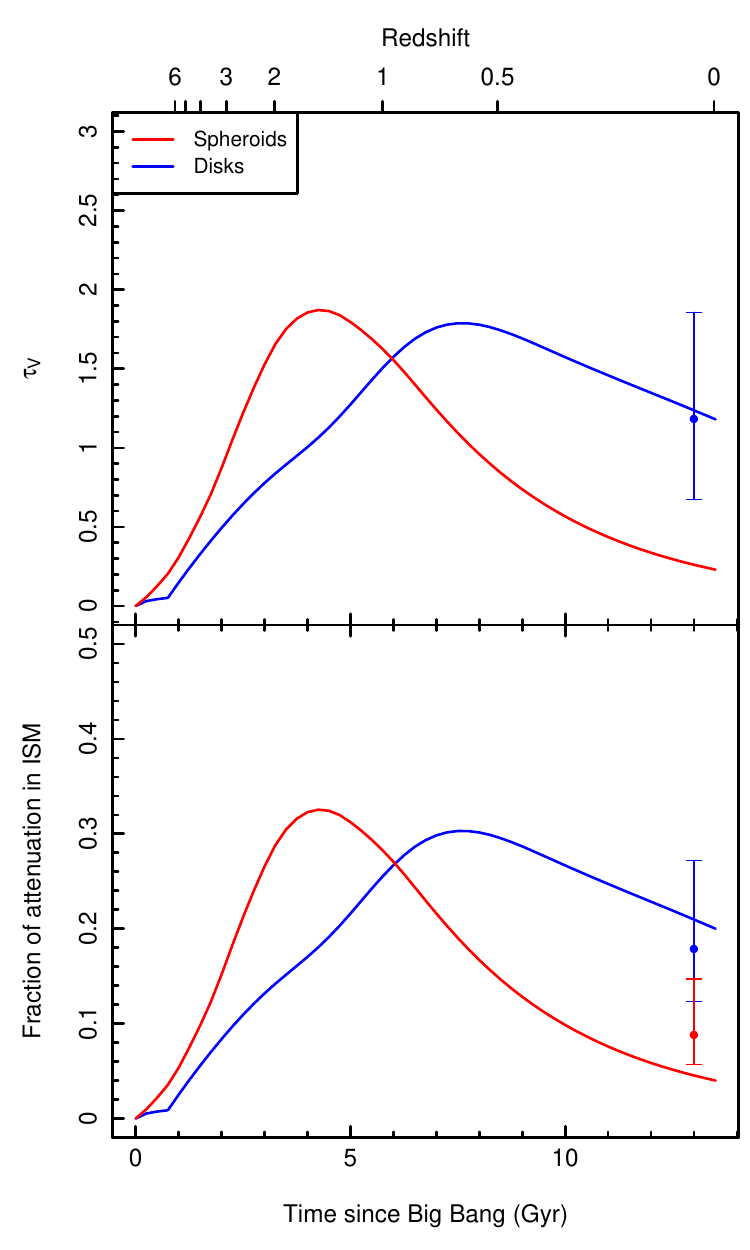}
\caption{The adopted $\tau_V$ (top) and $\mu$ (bottom) for spheroids (red) and discs (blue).}
\label{fig:tau}
\end{center}
\end{figure}

To model dust attenuation we use the \citet{charlot00} extinction model built in to \textsc{galaxev}. We set the $V$ band optical depth ($\tau_V$) as seen by stars in birth clouds to:

\begin{equation}
\tau_V(t)^a = X \exp\left(\frac{-t}{2.5}\right) \int_0^t~0.004~\mathrm{csfh}(t')~\exp\left(\frac{t'}{2.5}\right)~dt'
\end{equation}

where $t$ and $t'$ are ages of the Universe in Gyr, $X$ is the normalisation constant necessary to yield $\tau_V(13.5$~Gyr) = 1.2 for discs and 0.25 for spheroids, csfh($t$) is the CSFH for spheroids or discs and $a$ = 2 for discs and 1/0.65 for spheroids. The right hand side of the $\tau_V$ model is the dust mass density evolution from \citet{driver17}. This model is based on measurements of the dust mass density by \citet{menard12} and \citet{bethermin14}, and predicts that 0.4 per cent of mass forming into stars returns to the ISM as dust, with an exponential destruction with a characteristic time of 2.5~Gyr (see \citealt{driver17} for details). We choose the model involving exponential dust destruction -- the alternative model where some fraction of dust survives indefinitely is inconsistent with elliptical galaxies being largely dust free at $z = 0$. The values of $\tau_V(13.5$~Gyr) are chosen to yield a time evolution consistent with the median $\tau_V(13.5$~Gyr) value from the \citet{driver17} fits. We apply a similar prescription to the fraction of attenuation arising in the ISM $\mu$, setting $\mu$(13.5~Gyr) to be 20 and 4 per cent for discs and spheroids respectively. The chosen values for $\mu$ are guided by the median $\mu$($z < 0.06$) derived from spheroid and disc samples created by matching against the GAMA VisualMorphologyv03 catalogue\footnote{The spheroid value was derived from an elliptical sample defined by P\_EL\_DEBIASED $> 0.7$, while the disc value was derived from a subsample of Sd-Irr galaxies (HUBBLE\_TYPE\_CODE = 15). The corresponding values of $\tau_V$ for these samples are implausible.} Ultimately, the values of $\tau_V(13.5$~Gyr), $\mu$(13.5~Gyr) and the free parameter $a$ for discs are chosen to reproduce the ultraviolet, post-attenuation CSED for $0 < z < 1$ (especially at lower redshifts). The corresponding values regarding spheroids are less certain, but are chosen such that the resulting ultraviolet extinction is consistent with the empirical data. These functions and estimates are shown in Figure \ref{fig:tau}. The error bars shown represent the 16th and 84th percentile of the distribution as an indication only; we acknowledge that our empirically derived dust attenuation parameters are subject to large SED modelling errors due to limited sensitivity in the far-infrared at the current time.

We model the far-infrared emission with models from \citet{dale14} scaled up to the total energy attenuated. \citet{dale14} model the emission from dust exposed to 64 different heating intensities as described by the parameter $\alpha_\mathrm{sf}$, where model 1 ($\alpha_\mathrm{sf} = 0.0625$) has the most heating and model 64 ($\alpha_\mathrm{sf} = 4.0000$) the least. These models improve on the \citet{dale02} models used by \citet{driver13} by updating the mid-infrared observations on which the models are based and adding in emission from the dust torii of AGN. The main difference between these models and those used by \textsc{magphys} in the \citet{andrews16b} empirical CSED estimates is that the \citet{dale14} models produce less $70~\mu$m emission. This difference is immaterial given the lack of constraining data near this wavelength. As the \citet{dale14} templates are in arbitrary units, we scale to reflect the total attenuated energy implied by our stellar model and Equation 1 (integrated over 10~nm $< \lambda~<~8~\mu$m).

Initially we select the set of models with a zero AGN fraction. At all redshifts, we assume 15 per cent of the far-infrared radiation emitted originates from galaxies with warmer dust temperatures (such as ultra-luminous infrared galaxies) and 85 per cent from the broader infrared galaxy population. This assumption is reasonable, given $>90$ per cent of the normal infrared galaxy population and $>60$ per cent of the luminous infrared galaxy population have dust temperatures $< 35$~K out to $z = 0.5$ \citep{symeonidis13}. This ratio is slightly lower than the 25 per cent contribution from ultra-luminous infrared galaxies to the IGL at $140~\mu$m modelled by \citet{chary01}, however we find the lower ratio represents a better fit to the \citet{andrews16b} attenuated CSEDs. This is not definitive, given that the photometry underlying the \citet{andrews16b} far-infrared CSEDs is of relatively low depth and sensitivity; this also prevents computation of robust total infrared luminosity functions.

We select model 33 ($\alpha_\mathrm{sf} = 2.0625$) and 37 ($\alpha_\mathrm{sf} = 2.3125$) for infrared luminous and normal galaxy populations respectively at $z = 0$, the combination of which reproduces the \citet{andrews16b} far-infrared CSED between 100~$\mu$m and 500~$\mu$m for $z < 0.14$ well. At higher redshifts, we evolve the luminous galaxy model selection by selecting $\alpha_\mathrm{sf} = 1.3750 + 0.6875$ log(CSFH(13.5~Gyr)) / log(CSFH(t)), with $\alpha_\mathrm{sf}$ for normal galaxies being 0.500 greater than that for infrared luminous galaxies. This evolves the dust emission spectra towards a higher temperature on average -- from 25--32~K at $z = 0$ to 32--41~K at the peak of star formation at $z \sim 2$ -- in accordance with increased far-infrared emission and the $L_\mathrm{ir}$-$T_\mathrm{dust}$ relation \citep{symeonidis13}. These dust temperatures are also consistent with those observed by \citet{symeonidis13} over this redshift range. A weak dependence on the CSFH is also expected as increased heating from a higher CSFH is partially offset by increased dust masses, as noted by \citet{symeonidis13} and \citet{driver17}. This particular selection appears to reproduce the \citet{andrews16b} far-infared CSED well, however we caution that this estimate is based on a significant fraction of extrapolated flux. Furthermore, the \citet{symeonidis13} data probes galaxies with $L_\mathrm{ir} > 10^{10} L_\odot$ increasing to $10^{11.8} L_\odot$ at $z = 1$, and the normal infrared galaxy population out to $z = 0.3$, representing an incomplete picture of dust temperatures across the galaxy population out to $z = 1$.

\subsection{AGN modelling}

We represent emission from AGN in two classes -- obscured and unobscured. We derive an unobscured AGN composite spectrum from the sum of the SDSS composite quasar spectrum published by \citet{vandenberk01} and the 100 per cent AGN fraction \citet{dale14} model. To obtained an obscured AGN spectrum, we multiply the quasar composite spectrum by the integrated photon escape fraction as measured by \citet{andrews16b} for $0.82 < z < 0.99$ to represent dust attenuation and add the hottest \citet{dale14} model (number 1) to represent attenuation of AGN emission in the broader ISM. We scale the obscured and unobscured AGN spectra to yield a $g$ band rest frame luminosity equal to 2 and 1 times the eBOSS integrated $g$ band luminosity of \citet{palanque16} respectively (the \citet{palanque16} data traces the evolution of unobscured AGN) for an obscured to total AGN emission ratio of 67 per cent. Observations indicate a ratio of obscured to total AGN emission ranging between 41 to 77 per cent, with the majority of estimates toward the upper end \citep{richards06,buchner15}. This ignores the evolutionary biases traced by AGN (e.g. being more common in dust-free early type galaxies), but the impact of this assumption should be limited by the high photon escape fractions for these systems. AGN host galaxies were not included in the \citet{driver17} fits, and do not represent a significant addition to the \citet{andrews16b} CSED at any $z < 0.99$ -- \citet{andrews16b} find an extra 5--10 per cent contribution to the CSED from x-ray emitting AGN and their host galaxies except possibly in the ultraviolet as noted by \citet{driver16b}. 

\section{Model outputs}
\label{sec:csed}

\subsection{The unattenuated CSED}
\label{sec:ucsed}

\begin{figure*}
\begin{minipage}{7in}
\begin{center}
\includegraphics[width=0.325\linewidth]{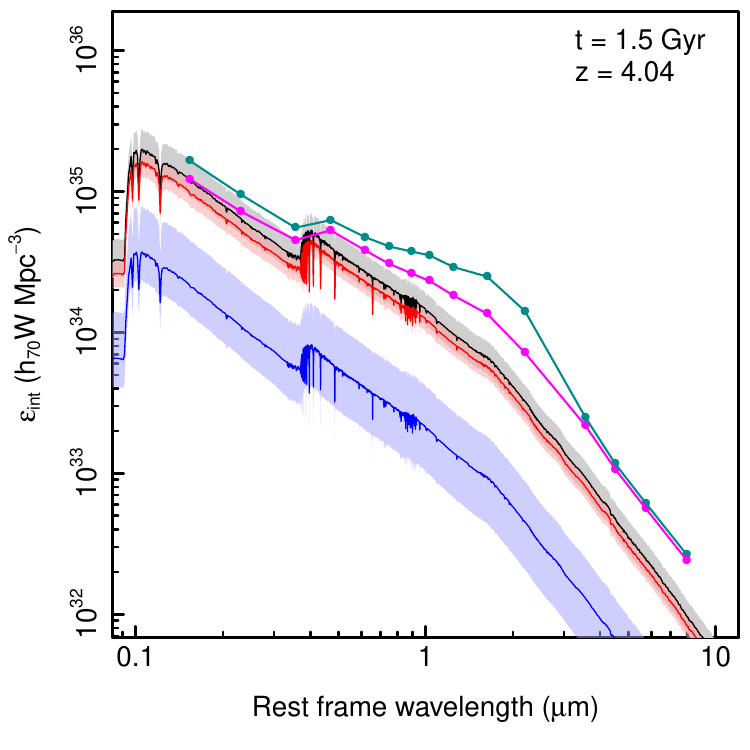}
\includegraphics[width=0.325\linewidth]{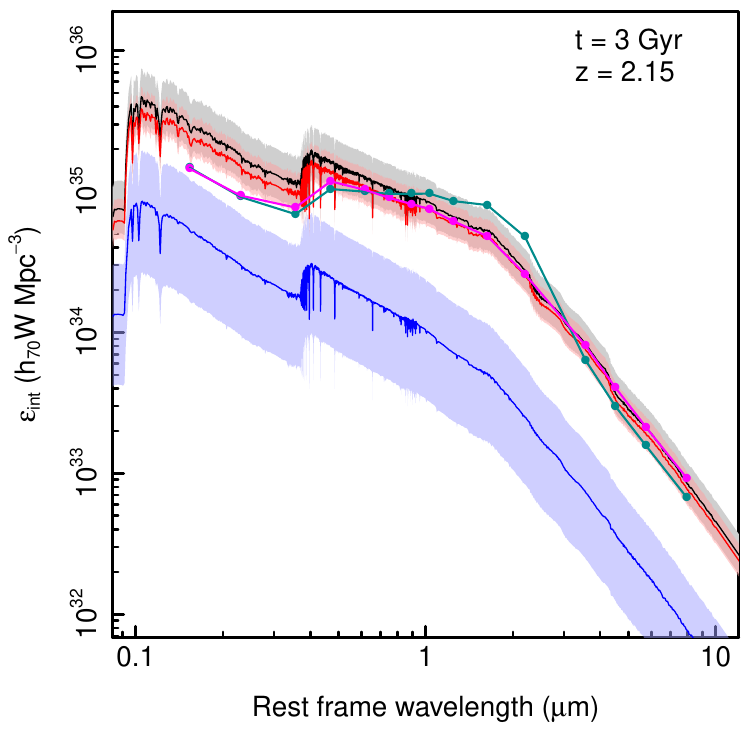}
\includegraphics[width=0.325\linewidth]{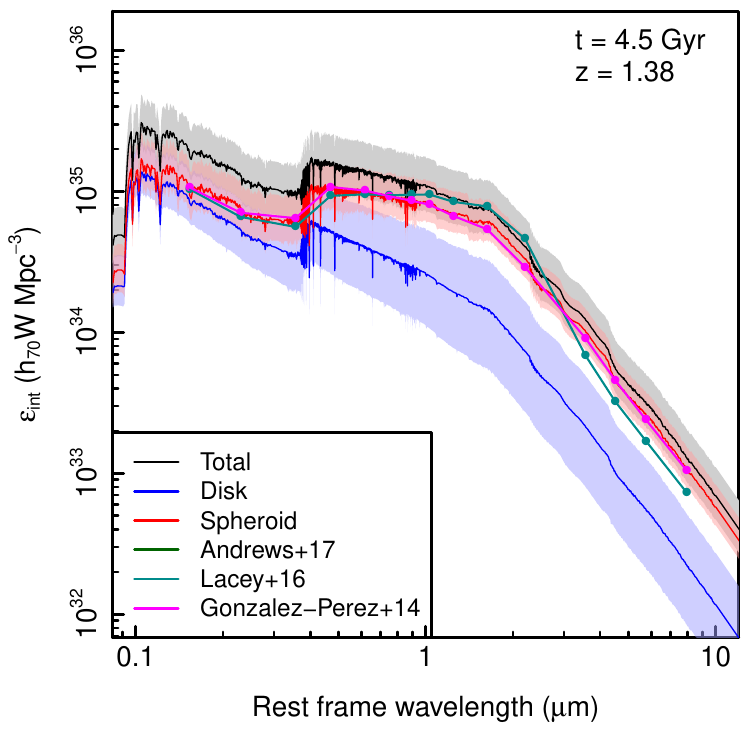} \\
\includegraphics[width=0.325\linewidth]{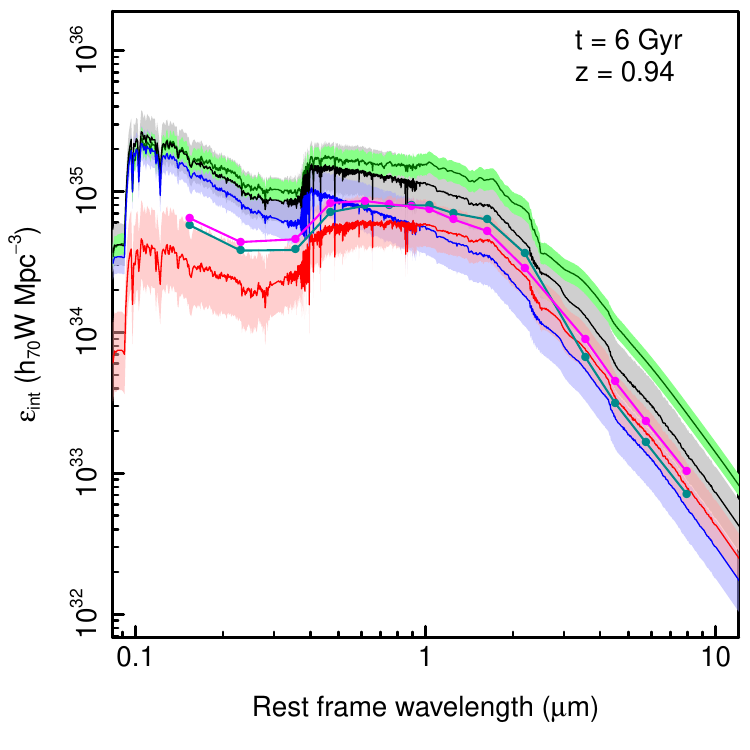}
\includegraphics[width=0.325\linewidth]{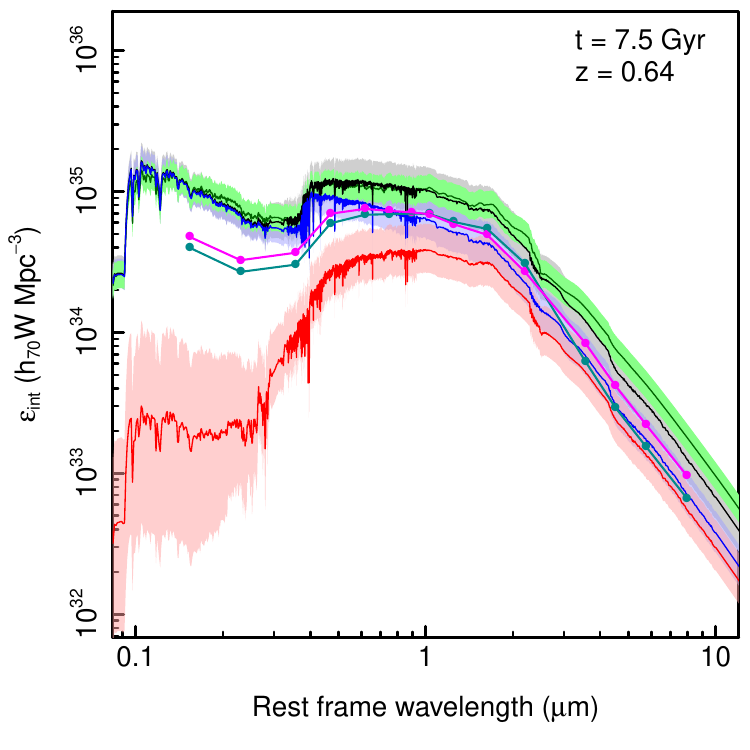}
\includegraphics[width=0.325\linewidth]{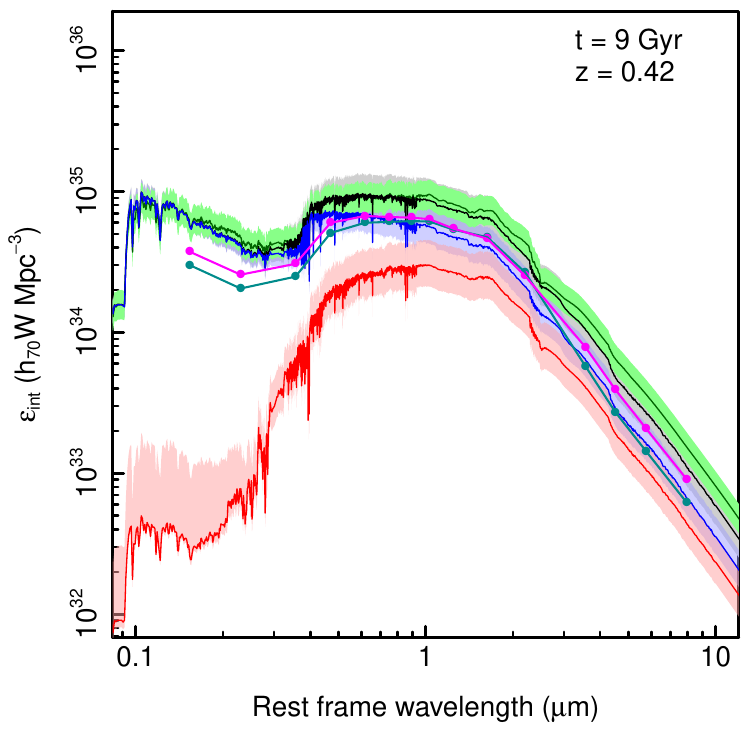} \\
\includegraphics[width=0.325\linewidth]{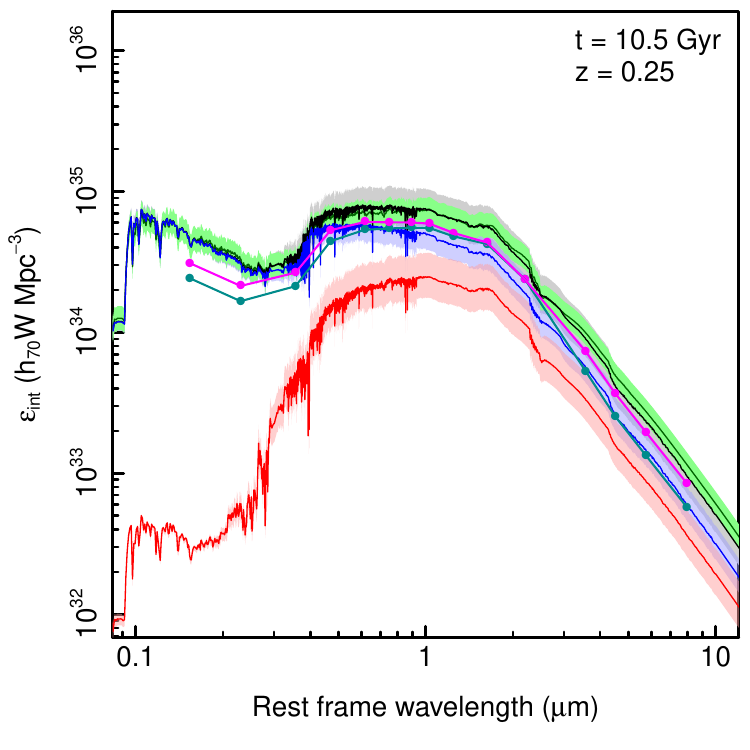}
\includegraphics[width=0.325\linewidth]{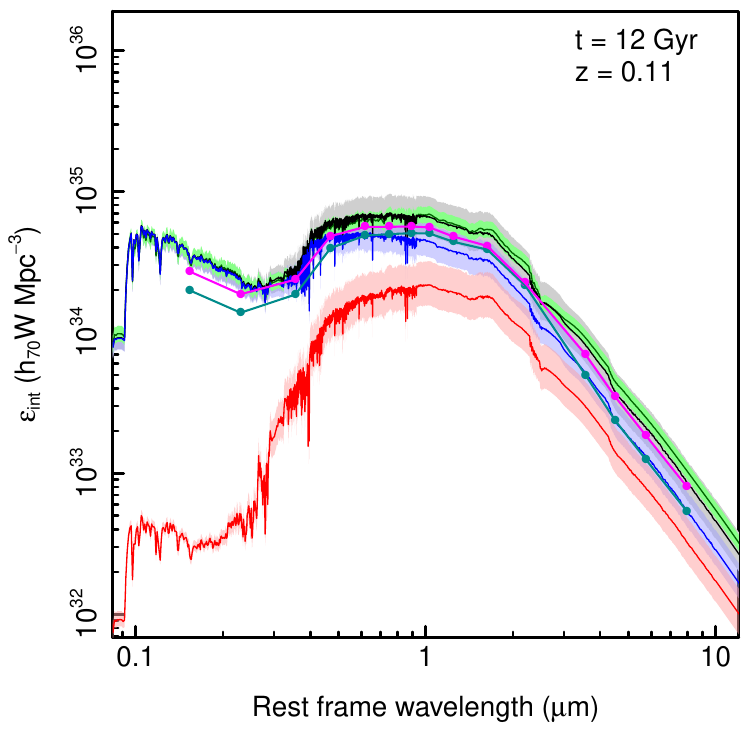}
\includegraphics[width=0.325\linewidth]{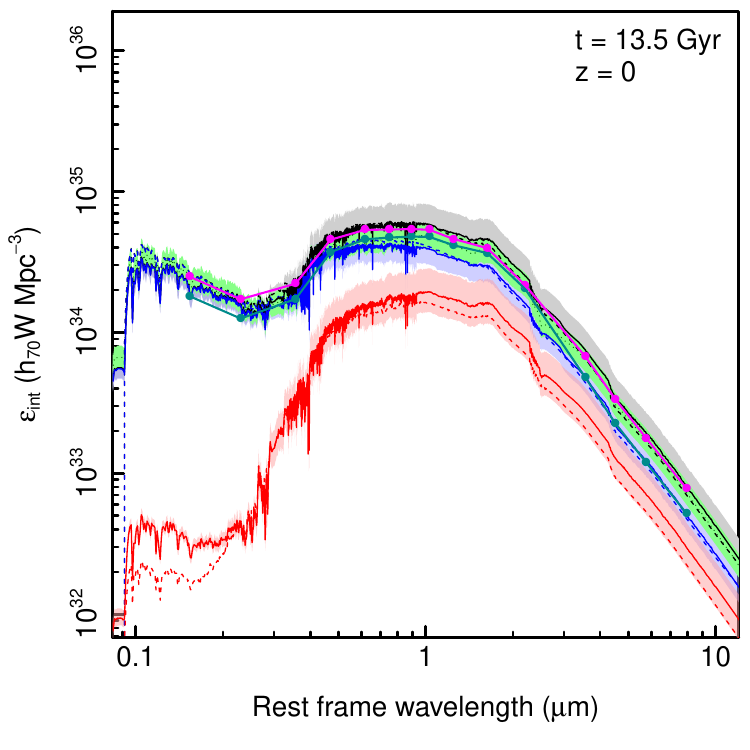}
\caption{The model unattenuated CSED for spheroids (red), discs (blue) and both combined (black), with shaded areas representing model uncertainty. The \citet{andrews16b} data is also given (green line/band), with the uncertainty range indicating the error in the normalisation of the CSED only. We also extrapolate this data to $z = 0$. Semi-analytic predictions from \textsc{galform} are shown in dark cyan \citep{lacey16} and pink \citep{gonzalezperez14}. An animated version of this Figure is available online as supporting information.}
\label{fig:mucsed}
\end{center}
\end{minipage}
\end{figure*}

Figure \ref{fig:mucsed} shows the predicted unattenuated CSED $\epsilon_\mathrm{int}$ (rest frame, in $\lambda f_\lambda$ units) from our phenomological model (black: total, blue: disc, red: spheroid) with shaded areas representing model uncertainty compared to the unattenuated \citet{andrews16b} estimates (for $0.02 < z < 0.99$) at the relevant timesteps (green band and line). While not shown in the Figure (as the data probes a period between 12.2~Gyr and 13.3~Gyr), the \citet{driver12} model lies within, but close to the lower bound of the uncertainty range of the CSED measurement. We also extrapolate the \citet{andrews16b} CSED estimates to $z = 0$ by rescaling the $0.02 < z < 0.08$ CSED by the expected decline in total integrated energy. The error range depicted in Figure \ref{fig:mucsed} for the \citet{andrews16b} data represents the error in the normalisation of the CSED only and does not incorporate any contribution from errors in SED modelling, incompleteness or photometry. 

From Figure \ref{fig:mucsed}, we can extract a number of key conclusions. The changeover from major mergers (spheroid formation) to cold gas accretion (disc formation) -- where emission from discs dominates in the rest frame ultraviolet -- occurs at $z \sim 1.2$, in line with the CSFH (see Figure \ref{fig:csfh}). Cold gas accretion primarily occurs in low-mass ($M_* <\sim 5 \times 10^{10} M_\odot$ for a $0 < z < 0.5$ GAMA sample) galaxies, with merger accretion continuing in high-mass galaxies \citep{robotham14}. The total energy output of the Universe reaches a maximum of $5.0_{-1.6}^{+2.7} \times 10^{35}~h_{70}$~W~Mpc$^{-3}$ at $z \approx 2.1$, while energy output from discs reaches a maximum of just $2.7_{-0.7}^{+1.0} \times 10^{35}~h_{70}$~W~Mpc$^{-3}$ at $z \approx 0.8$. To date, just above half (50.6 per cent) of the energy generated by the Universe was generated by objects now residing in discs, with the other half being generated by material now residing in spheroids. For reference, 50 per cent of the stellar mass today resides in spheroids, with the other 50 per cent in discs \citep{moffett16b}.

Recall that the model is calibrated on the CSFH, and therefore is designed to reproduce the (unattenuated) ultraviolet emission. However, the \citet{andrews16b} empirical CSEDs have a number of caveats which may become reflected in the model. Firstly, the empirical CSEDs may suffer from incompleteness due to Malmquist bias, resulting in the omission of low mass, blue, star forming systems. \citet{andrews16b} estimate that this incompleteness causes on the order of 20 per cent of the ultraviolet flux to be missed. Secondly, the \citet{wright16a} catalogue, on which the empirical CSEDs are based, measures ultraviolet flux in an optically defined aperture convolved with the \textit{GALEX} point spread function. This approach may miss additional, extended ultraviolet emission from disc galaxies \citep{gildepaz05,thilker07} for $z < 0.45$.

Our model reproduces the unattenuated CSED in the near-infrared well, except arguably at $z \sim 0.9$, where we see a slight deficit in the model. Uncertainties in the near-infrared may be the result of a number of factors:

\begin{itemize}
\item Uncertainties in modelling thermally pulsating asymptotic giant branch stars. We have tried to control for this effect, as both the phenomological model and CSED estimates are based on the \citet{bc03} stellar libraries. 
\item Uncertainties and imprecisions in modelling gas-phase metallicity. However, Figure \ref{fig:metallicity_spectra} shows these have a much greater effect in the ultraviolet and optical shortward of the 4000~\AA~break and opposite effects either side of about 1~$\mu$m. This would suggest that the lack of metallicity interpolation is not the dominant source of modelling error.
\item The photometric and spectroscopic data underlying the \citet{andrews16b} measurements use an observed frame $r$ or $i^+$ band selection. At the high-redshift ends of GAMA and G10/COSMOS, this is equivalent to $u$ or $g$ in the rest frame. It is, hence, likely that some quiescent or very dusty galaxies are excluded.
\item The \citet{driver13} model applies a 25 per cent downward adjustment to the CSFH for spheroids, resulting in a spheroid to disc mass ratio at the current epoch of $\sim\frac{2}{3}$. This renormalisation is not necessary in our model -- see Section \ref{sec:stellarmass}.
\end{itemize}

Overall, our phenomological model appears to reproduce the \citet{andrews16b} unattenuated CSED for $z < 1$ extremely well. The black model line lies close to, or within the green measurement band (which takes into account uncertainty in the normalisation of the CSED only, excluding the biases noted above and in \citealt{andrews16b}) at all redshifts -- with just a minor discrepancy in the near-infrared CSED in the $0.82 < z < 0.99$ bin. The model uncertainty at optical wavelengths and low redshifts is mostly due to uncertainty in the CSFH at high redshifts. An animated version of Figure \ref{fig:mucsed} is available online as supporting information.

\subsection{The attenuated CSED}
\label{sec:acsed}

\begin{figure*}
\begin{minipage}{7in}
\begin{center}
\includegraphics[width=0.325\linewidth]{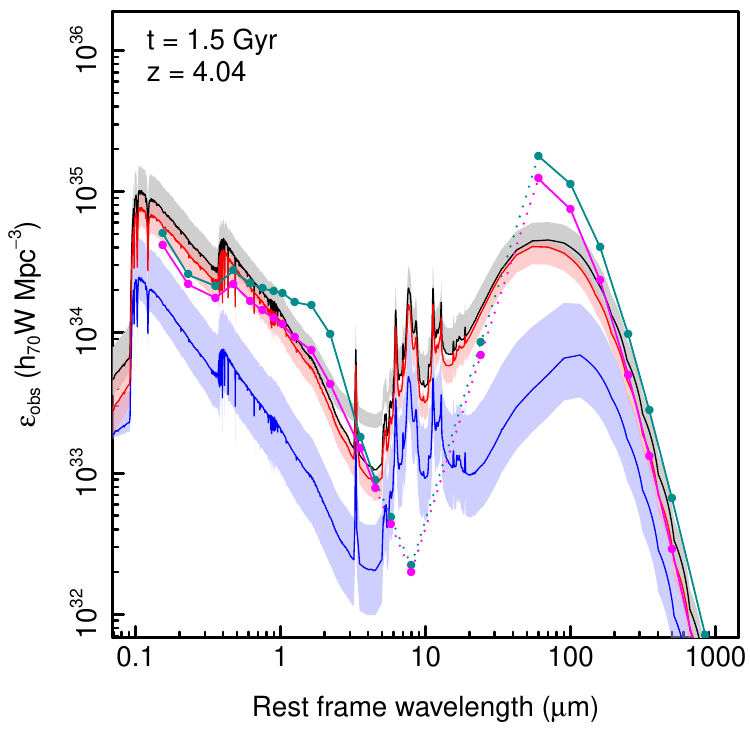}
\includegraphics[width=0.325\linewidth]{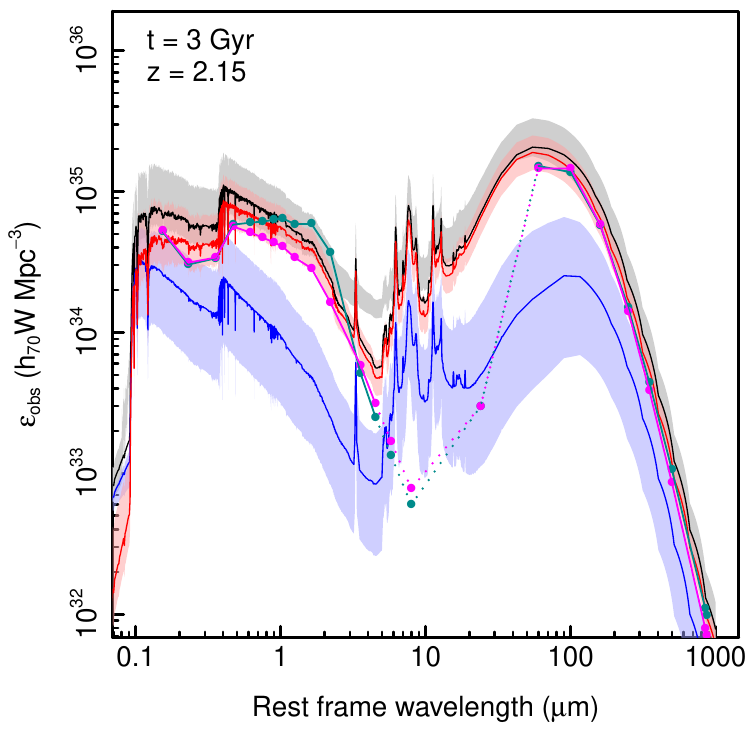}
\includegraphics[width=0.325\linewidth]{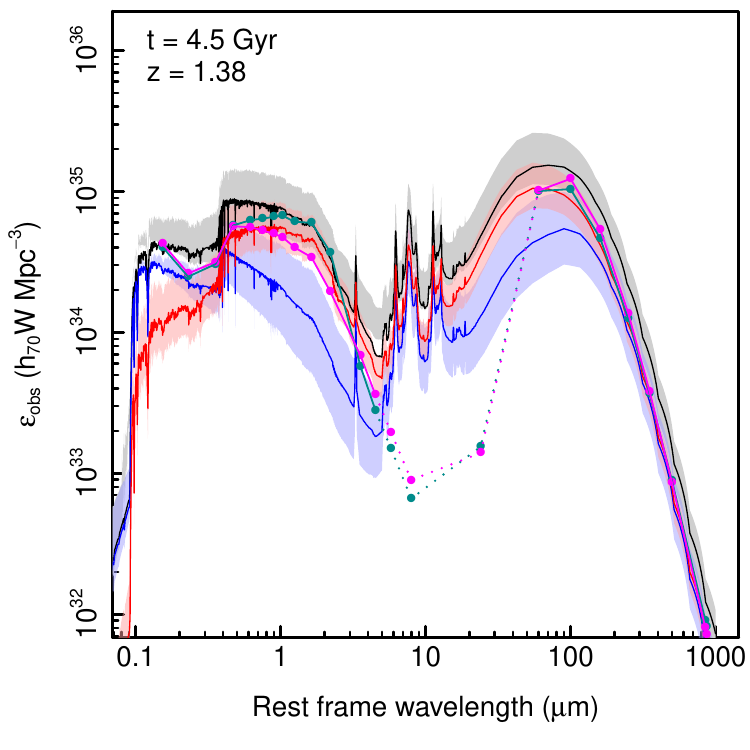} \\
\includegraphics[width=0.325\linewidth]{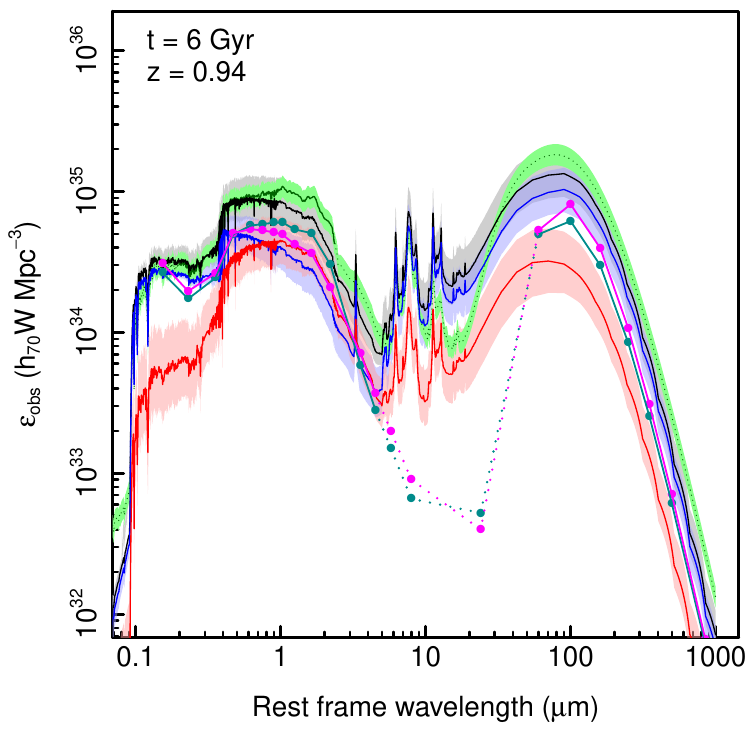}
\includegraphics[width=0.325\linewidth]{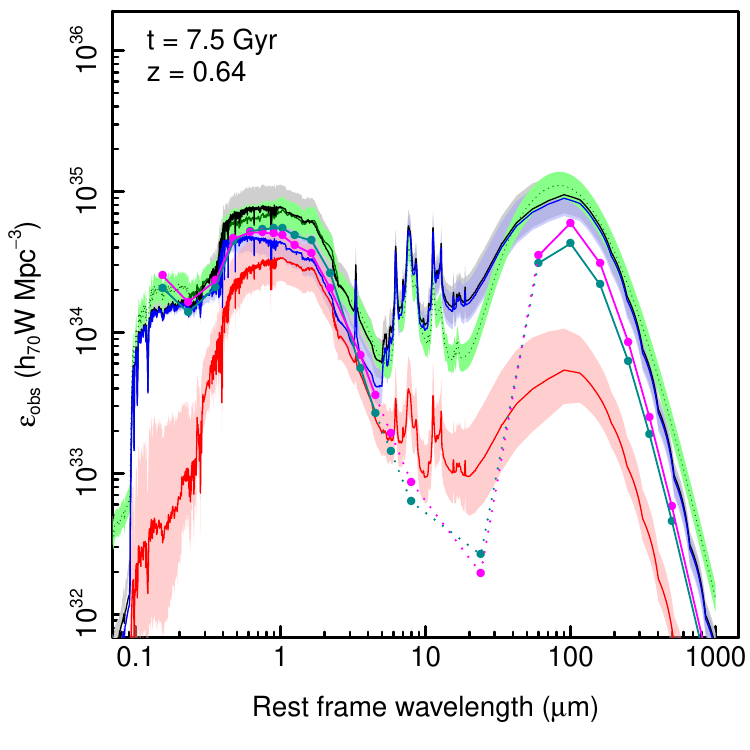}
\includegraphics[width=0.325\linewidth]{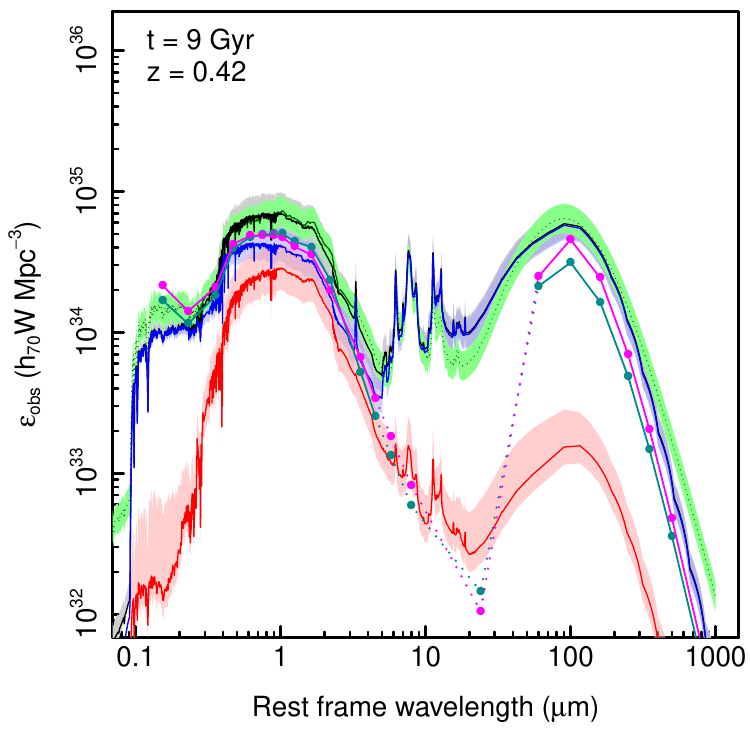} \\
\includegraphics[width=0.325\linewidth]{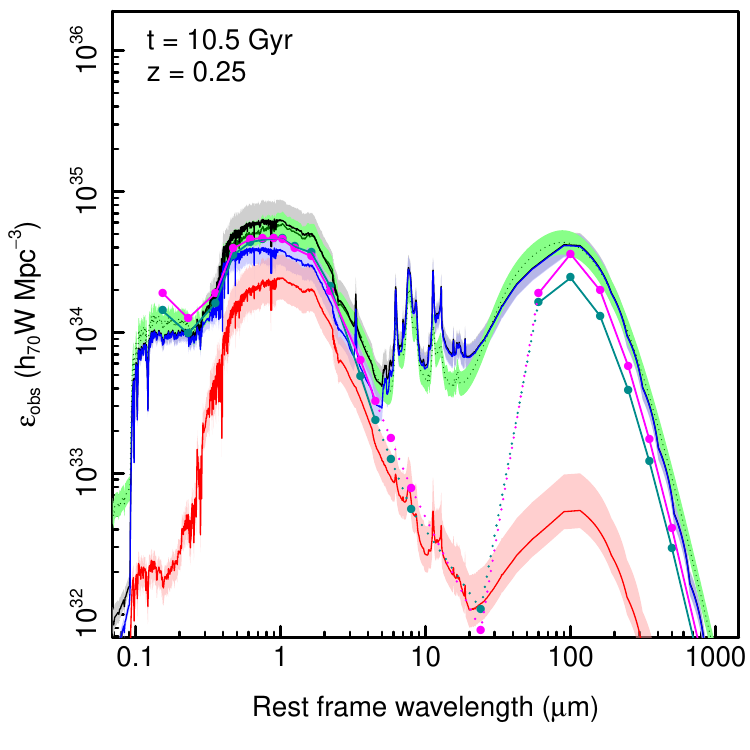}
\includegraphics[width=0.325\linewidth]{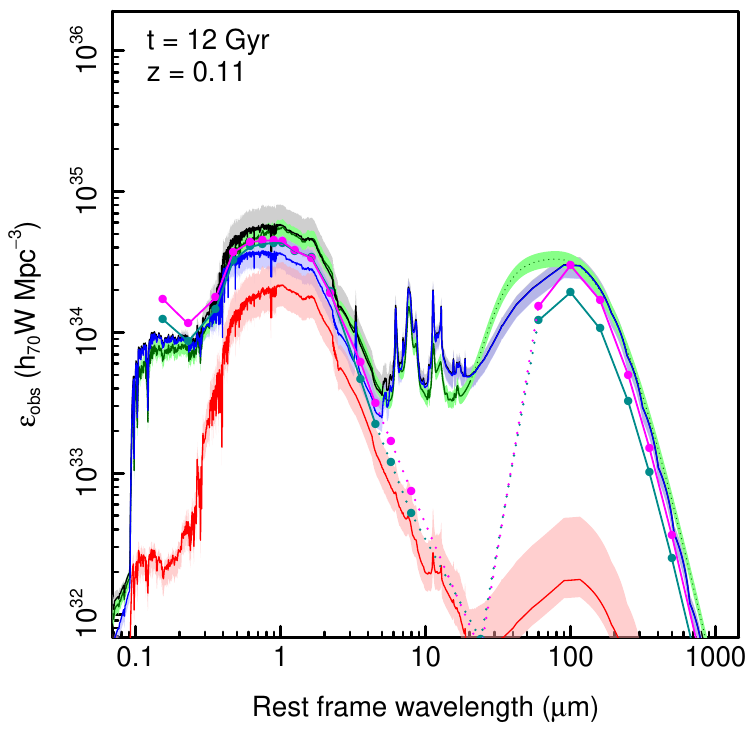}
\includegraphics[width=0.325\linewidth]{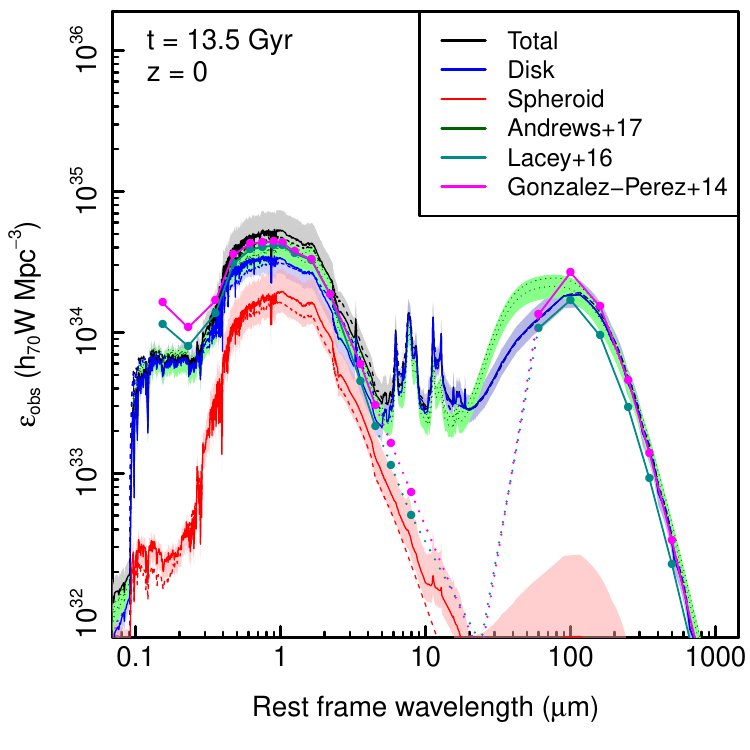}
\caption{The model attenuated CSED with labelling equivalent to Figure \ref{fig:mucsed}. Wavelength intervals where the \citet{andrews16b} CSEDs may be deemed less reliable due to the lack of underlying photometric data and where semi-analytic predictions from \textsc{galform} omit flux from warm dust are denoted with a dotted line. An animated version of this Figure is available online as supporting information.}
\label{fig:mfcsed}
\end{center}
\end{minipage}
\end{figure*}

We now add in our dust prescription as described in Section \ref{sec:dust} to determine the attenuated (as observed) CSEDs. Figure \ref{fig:mfcsed} shows the resulting attenuated CSED $\epsilon_\mathrm{obs}$, now with far-infrared emission at various timesteps with the \citet{andrews16b} attenuated CSED data at the relevant timesteps. The \citet{andrews16b} data have been extrapolated to $z = 0$. The \citet{driver12} model, being relevant for $0.013 < z < 0.1$ and thus not shown above, generally lies within the uncertainty range of the empirical CSED estimates. Again, the error range in the empirical CSEDs reflects the error in the normalisation of the CSED only, excluding uncertainties in SED modelling and the underlying photometry. 

Both optical and far-infrared emission reach a maximum of $2.2_{-0.8}^{+1.4}$ and $3.7_{-1.1}^{1.7} \times 10^{35}~h_{70}$~W~Mpc$^{-3}$ at $z \approx 2.1$.

The most noticeable disagreement between the phenomological model and the CSED data occurs at around 70~$\mu$m. Empirically, \textsc{magphys} is unable to constrain the warm dust peak resulting from the lack of sufficiently deep 70~$\mu$m data in both the GAMA and G10/COSMOS data sets. The choice of different dust emission templates to model the far-infrared emission is also important --- \citet{driver12} use the average of \citet{dale02} models 34 and 40, which results in a small warm dust peak that is slightly below our model curve. Ultimately, the far-infrared imaging data underlying the SED fits is of poorer quality in terms of resolution and sensitivity -- the CSED estimates are only precise to a factor of about two at low redshifts, increasing to a factor of five beyond $z = 0.45$. Estimates in G10/COSMOS are based on a significant fraction of extrapolated flux, while in GAMA the low detection rate undermines the reliability of the corresponding CSED (see Figure 7 in \citealt{andrews16b}). We denote wavelength intervals where the CSED estimates may be unreliable with a dashed (as opposed to solid) line. For this reason, a rigourous optimisation against the \citet{andrews16b} CSED estimates between $100 \mu$m and $500~\mu$m is not possible.

Otherwise, our phenomological model again reproduces the attenuated CSED well at all redshifts -- lying within or close to the green error bounds of the normalisation of the CSED only (which excludes SED modelling error, incompleteness and bias) with two exceptions, the first being the $z \sim 0.91$ shortfall in the near-infrared mentioned previously. Secondly, the model overpredicts flux at 20~$\mu$m. This is a consequence of differing prescriptions of warm dust used by the empirical CSED measurements and the model predictions, coupled with the difficulty of correctly modelling emission from polycyclic aromatic hydrocarbons and the incompleteness of the observational data at this wavelength. An animated version of Figure \ref{fig:mfcsed} is available online as supporting information.

\section{Semi-analytic modelling with \textsc{galform}}
\label{sec:sams}

The phenomological model described in Section \ref{sec:toymodel} is a simplistic description built explicitly to model the CSED, providing an instant CSED prediction at any redshift to help interpret the extragalactic background light and photon escape fraction (see below). While it has predictive power outside this domain --- one can infer the cosmic stellar and dust mass densities --- many phenomena are neglected (for example, variations with environment). Semi-analytic models are able to provide a broader understanding, but at the expense of additional complexity.

Here, we use two recent versions of the \textsc{galform} semi-analytic model \citep{gonzalezperez14,lacey16} to derive predictions of the CSED across cosmic time. We derive predictions for the CSED for snapshots close to the mean redshift in the \citet{andrews16b} redshift bins and at $z = 1.173$, 1.504, 2.070, 2.422, 3.060, 3.576, 4.179, 4.520, 5.724 and 6.197. For each snapshot, we compute the pre- and post-attenuated rest frame luminosity distributions as they would be observed through \textit{GALEX}, SDSS, UK Infrared Telescope, IRAC bands and at 24, 60, 100, 160, 250, 350, 500, 850 and 870~$\mu$m. We then multiply by the respective luminosities and sum to arrive at the corresponding luminosity density for each band. The resulting predicted (rest frame) unattenuated and attenuated CSEDs are shown in Figure \ref{fig:mucsed} and \ref{fig:mfcsed} respectively -- the \citet{lacey16} is denoted with dark cyan curves, while the \citet{gonzalezperez14} model is denoted with pink curves.

The two \textsc{galform} flavours presented here were calibrated to fit the $B_j$ and $K$ band luminosity functions at $z = 0$, and the evolution of the $K$ band luminosity function out to $z \sim 3$. In addition, the \citet{lacey16} model paid close attention to number counts and redshift distributions of sources observed by \textit{Herschel} and at $850~\mu$m. Therefore, the comparisons we derive here (i.e. the IGL and evolution of the CSED) are mostly independent tests that can offer valuable insight into how the models can be improved.

These instances of \textsc{galform} split star formation into two modes in a similar manner as our phenomological model -- a quiescent mode, driven by cold gas accretion onto discs and a starburst mode, where galaxy mergers and disc instabilities transfer disc gas into spheroids triggering a starburst. The primary difference between the two versions of \textsc{galform} is the choice of IMF. The \citet{lacey16} version employs different IMFs for spheroids and discs -- \citet{kennicutt83} for disc star formation and a custom, top-heavy IMF -- a power law with a tunable slope between 0 and 1 -- for spheroid star formation. In contrast, the \citet{gonzalezperez14} version assumes a Universal \citet{kennicutt83} IMF. \textsc{galform} predicts an earlier transition from spheroid to disc formation than the phenomological model -- $z \sim 3$ compared to $z \sim 1.2$ (see \citealt{lacey16}, Figure 26).

Mock photometry is computed in the \citet{lacey16} model using the \citet{maraston05} stellar population synthesis codes, while the \citet{gonzalezperez14} model uses the \citet{bc03} libraries. The \citet{maraston05} libraries track fuel consumption, in contrast to the isochrone analysis employed by Bruzual \& Charlot. The modelling of thermally pulsating asymptotic giant branch stars is still controversial, with the Maraston libraries arguably overestimating the near-infrared emission (see e.g. \citealt{maraston05,maraston06,bruzual07,conroy10,bruzual13,noel13,capozzi16}). The use of the Maraston stellar libraries is the most likely cause of the near-infrared enhancement of the \citet{lacey16} model over both the phenomological model and the \citet{gonzalezperez14} model at $z > 1$ in both the unattenuated and attenuated CSEDs (see Figures \ref{fig:mucsed} and \ref{fig:mfcsed}).

\textsc{galform} can also compute dust attenuation via radiative transfer through a two-phase medium consisting of molecular clouds and the diffuse ISM, with dust emission described using a modified blackbody spectrum \citep{lacey16}. \textsc{galform} does not model the mid-infrared emission as it does not incorporate polycyclic aromatic hydrocarbons \citep{cowley17}. As a consequence, the predicted attenuated CSED is unreliable between 8~$\mu$m and 70~$\mu$m rest frame. This region is denoted with a dotted line in Figure \ref{fig:mfcsed}.

Both \textsc{galform} models underpredict the cosmic star formation history for $z < 3$ \citep{mitchell14,guo16,lacey16} compared to literature estimates (e.g. \citealt{madau14}). The predictions, however, agree with the lower cosmic star formation history derived by \citet{driver17} (on which the empirical CSED estimates are based). However, \textsc{galform} produces a fainter unattenuated CSED than our empirical results. When adjusted for this offset, both semi-analytic models produce a shape of the unattenuated CSED that is very similar to our CSED estimates. Beyond $z = 3$, the \textsc{galform} predictions of the cosmic star formation history show a better agreement with observations. Far-infrared emission grows faster than optical emission in both iterations of \textsc{galform} at very high redshifts. This originates from obscured star formation in ultra-luminous infrared galaxies, which contribute a much greater portion of the total far-infrared luminosity at higher redshifts \citep{lagos14}. This effect is not accounted for in the phenomological model.

When adjusted for the offset in the CSED normalisation, both \textsc{galform} models reproduce the shape of the optical and near-infrared attenuated CSED well out to $z = 1$ (Figure \ref{fig:mfcsed}). Na\"{i}vely, the \citet{gonzalezperez14} model yields a better fit in the far-infrared. However, when adjusted, both models overpredict the ultraviolet CSED, with the \citet{lacey16} model being closer to the empirical data. We suspect the underpredicted dust attenuation is a result of overpredicted galaxy sizes -- \citet{merson16} show that \textsc{galform} is able to derive reasonable predictions for dust attenuation when the predicted half-mass radii are also plausible. However, galaxies with overpredicted sizes have much less attenuation than expected, as the diffuse dust component has a lower surface density.

In summary, both iterations of \textsc{galform} underpredict the cosmic star formation history for $z < 3$ and thus the normalisation of the unattenuated and attenuated CSEDs. Both semi-analytic models are able to reproduce the shape of the unattenuated and attenuated CSEDs well out to $z < 1$, with the exception that both models seem to underpredict the ultraviolet attenuation over $0 < z < 1$. In the absence of higher redshift CSED estimates, we look to constraints on the extragalactic background light and cosmic optical and infrared backgrounds.

\section{Application of the model}
\label{sec:results}

Having developed a model which replicates the unattenuated and attenuated CSEDs for $z < 1$, we can now explore predictions of related quantities, such as the photon escape fraction, IGL, and stellar and dust mass densities compared to observations. 

\subsection{The integrated photon escape fraction}
\label{sec:ipef}

\begin{figure*}
\begin{minipage}{7in}
\begin{center}	
\includegraphics[width=0.325\linewidth]{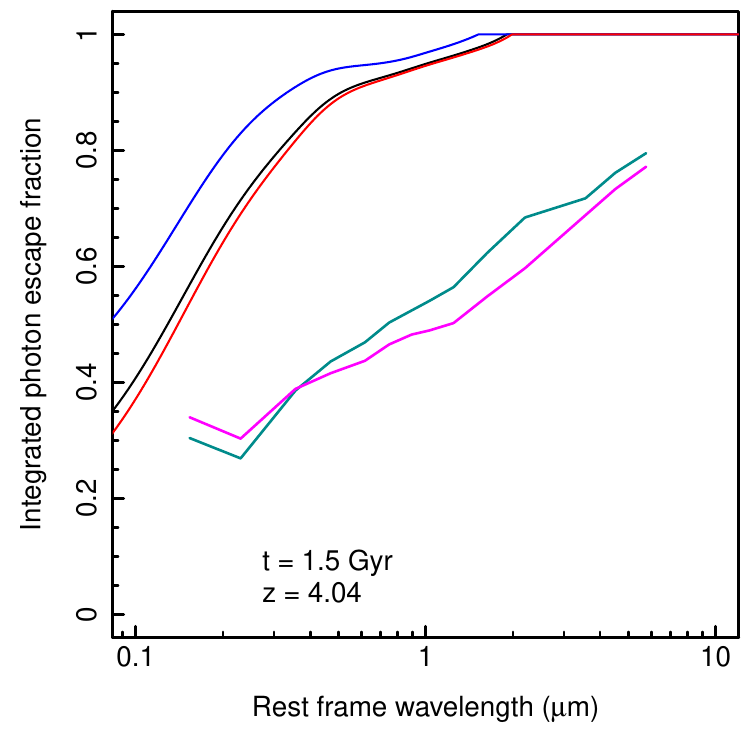}
\includegraphics[width=0.325\linewidth]{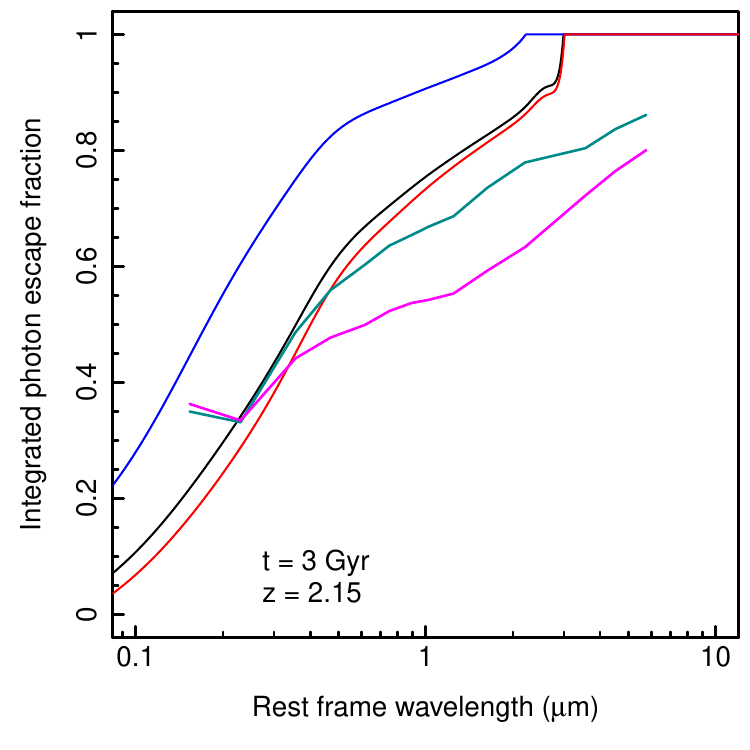}
\includegraphics[width=0.325\linewidth]{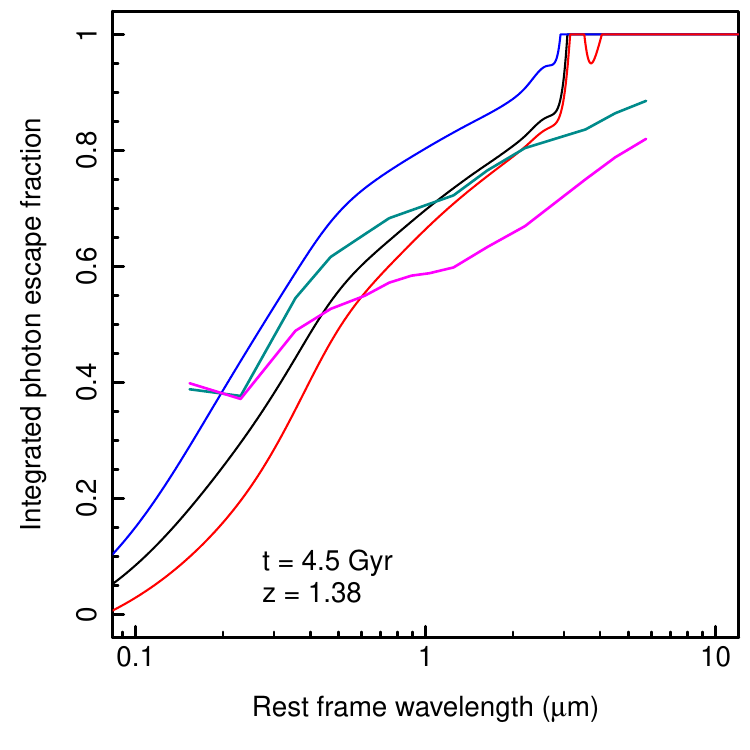} \\
\includegraphics[width=0.325\linewidth]{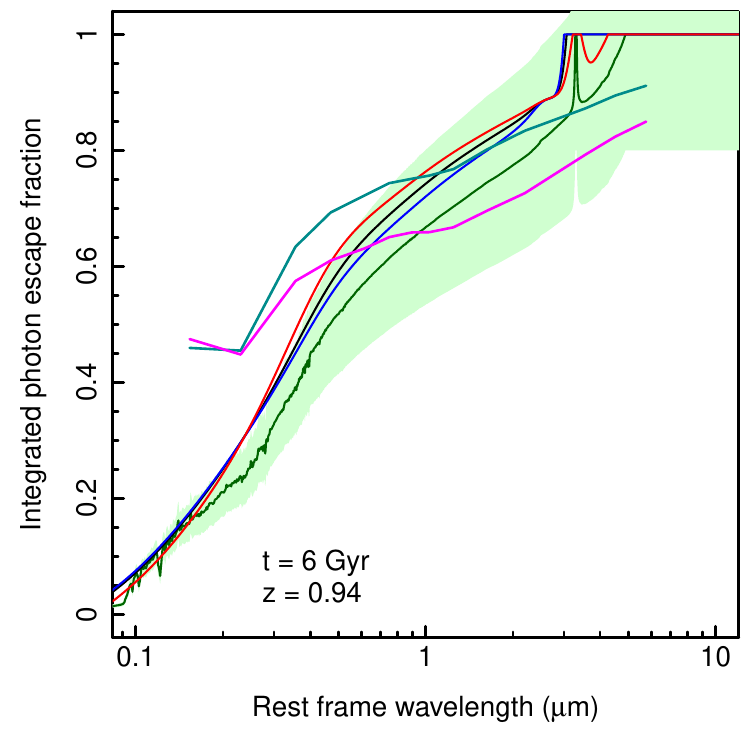}
\includegraphics[width=0.325\linewidth]{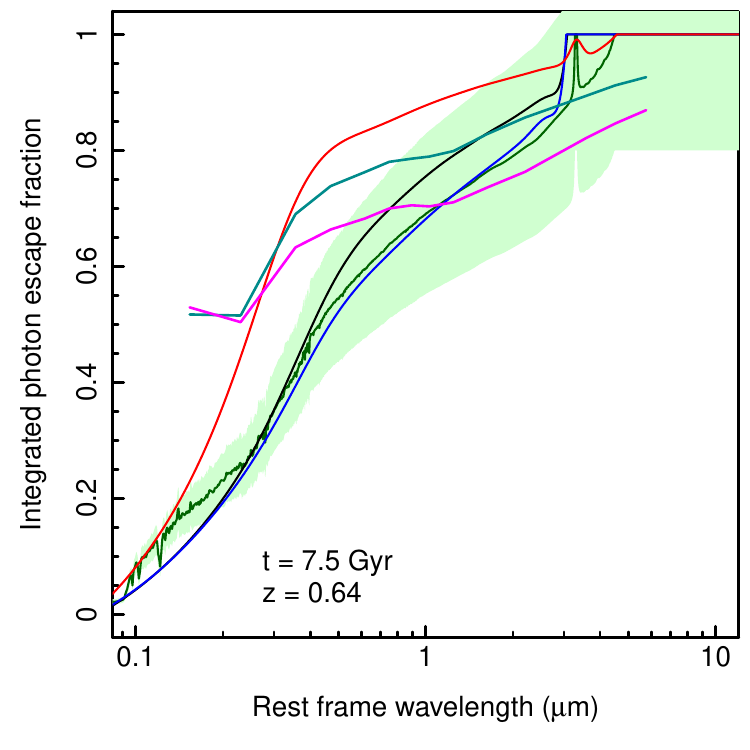}
\includegraphics[width=0.325\linewidth]{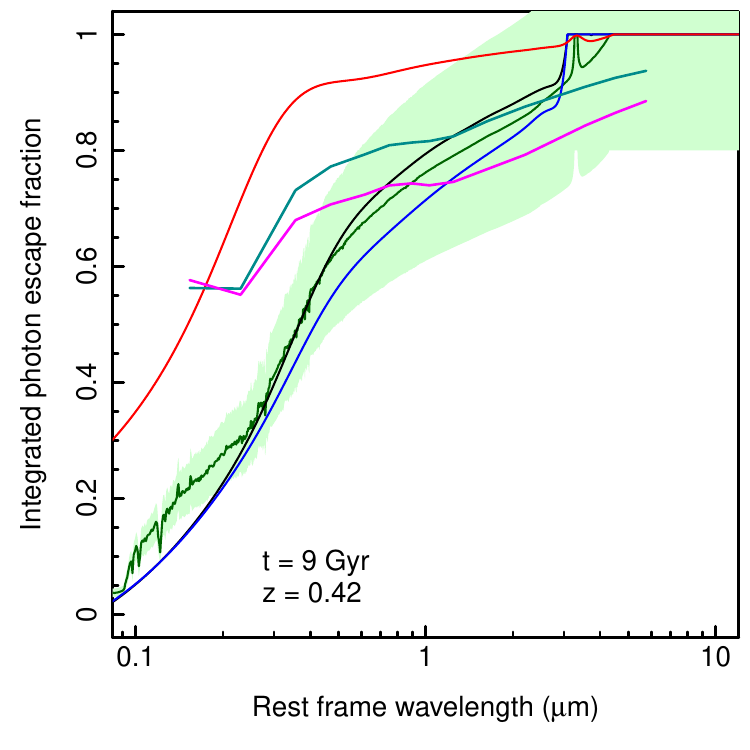} \\
\includegraphics[width=0.325\linewidth]{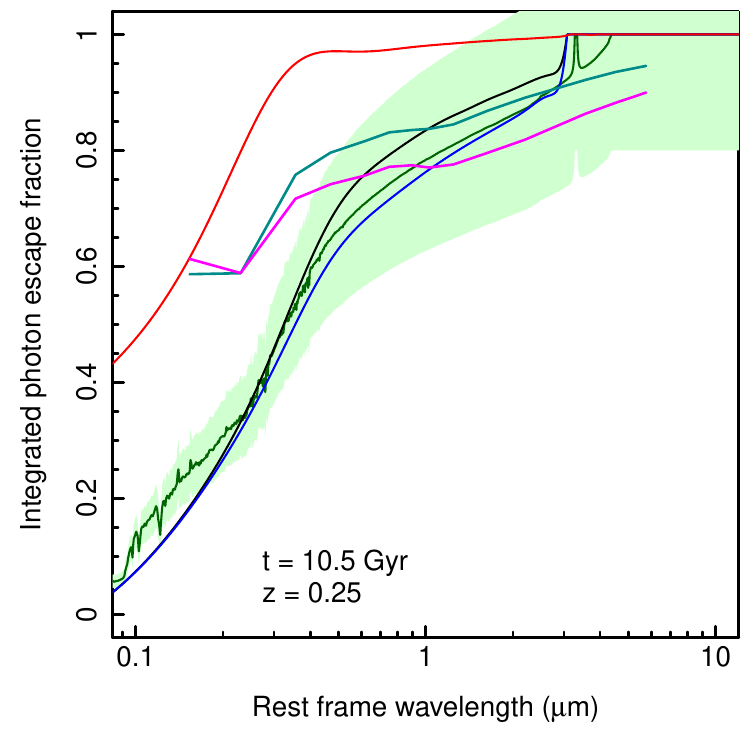}
\includegraphics[width=0.325\linewidth]{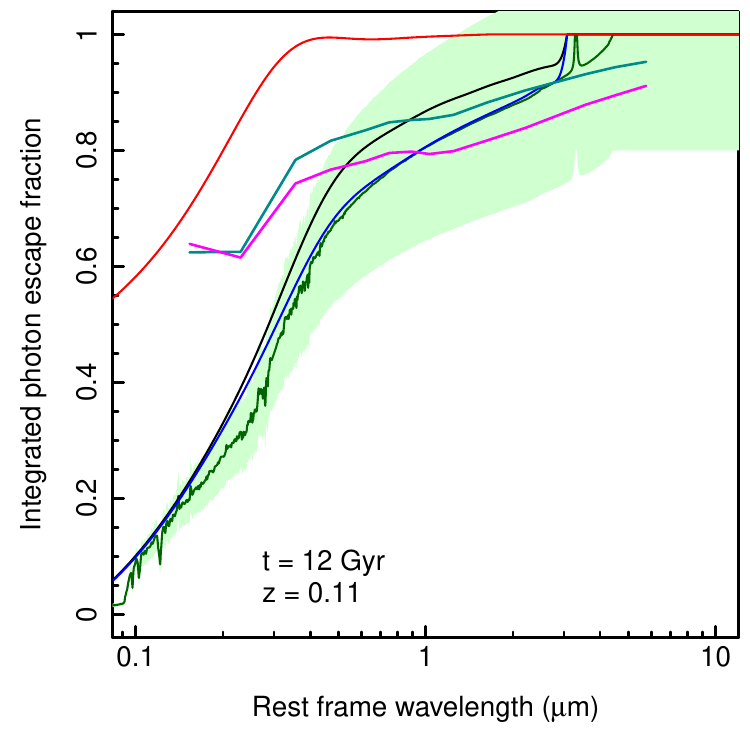}
\includegraphics[width=0.325\linewidth]{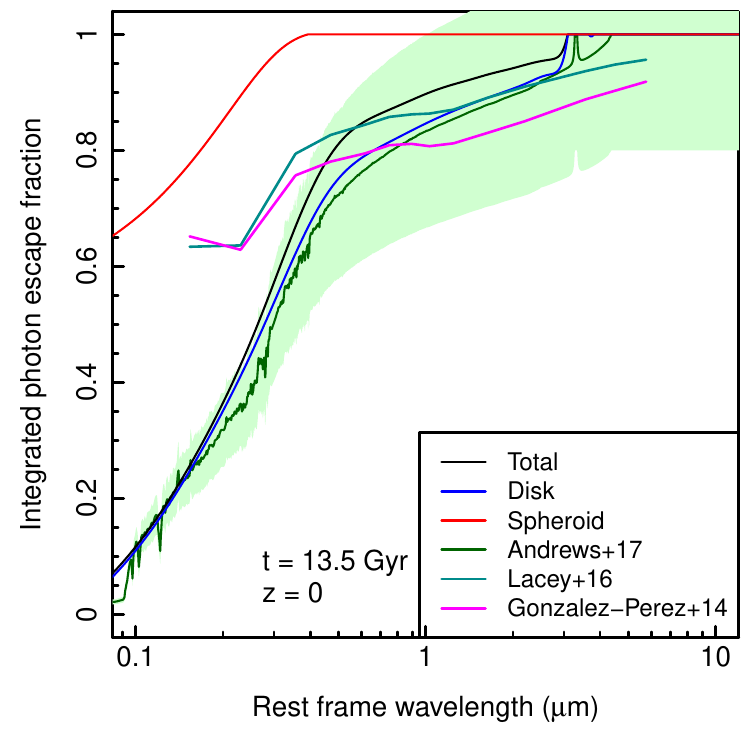}
\caption{The model IPEF for spheroids (red), discs (blue) and both combined (black) with semi-analytic predictions from \textsc{galform} (dark cyan, pink). The \citet{andrews16b} data is also given in green.}
\label{fig:mipef}
\end{center}
\end{minipage}
\end{figure*}

The integrated photon escape fraction (IPEF) represents the attenuated CSED divided by the unattenuated CSED. It is a simplistic but useful representation of the effects of dust attenuation. It is particularly useful for correcting ionising radiation pervading the ISM and/or determining unattenuated star formation rates. The predominant source of error in the IPEF arises from SED modelling and incompleteness -- the uncertain normalisation due to cosmic variance, sampling and the use of photometric redshifts is divided out. The uncertainty arising from the CSFH also cancels.

Figure \ref{fig:mipef} shows IPEFs as a function of redshift for both our model (smoothed with a spline interpolation) and the \citet{andrews16b} data (green shaded band). The disc IPEF reproduces the \citet{andrews16b} IPEFs well, especially at lower redshifts. The overall IPEF reaches a minimum longwards of $\lambda = 700$~nm at $z \sim 1.7$. The ultraviolet IPEF has two minima -- one at $z \approx 1.7$, the other at $z \approx 0.6$. Spheroids become more transparent than discs across the electromagnetic spectrum at $z \sim 0.9$. Our methodology assumes dust emission is negligible. The spike in the photon escape fraction at approximately 3~$\mu$m is a consequence of the $3.3~\mu$m polycyclic aromatic hydrocarbon emission feature and should not be regarded as an estimate of the photon escape fraction at that wavelength.

As noted before and in \citet{gonzalezperez17}, both iterations of \textsc{galform} underpredict the amount of dust attenuation for $z < 1$ compared to the \citet{andrews16b} empirical CSED estimates. The lack of a prescription for warm dust is the most likely cause for the shortfall in the mid-infrared. 

\begin{figure}
\begin{center}
\includegraphics[width=0.99\linewidth]{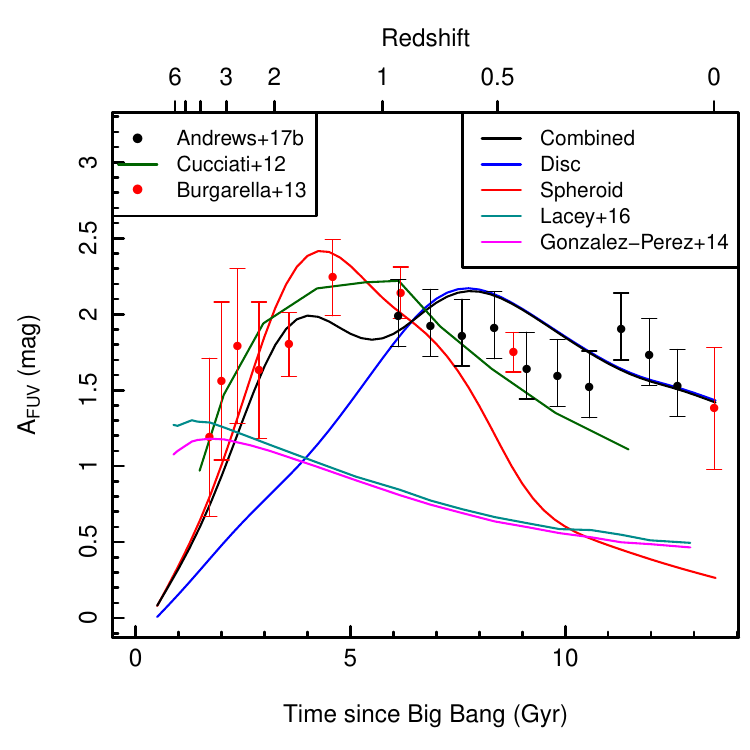}
\caption{Rest frame $A_\mathrm{FUV}$ as a function of redshift from the phenomological model (blue: disc, red: spheroids, black: weighted average by the cosmic star formation rate in each component), \textsc{galform}, the \citet{andrews16b} estimates, \citet{cucciati12} and \citet{burgarella13}.}
\label{fig:fuvattenuation}
\end{center}
\end{figure}

One commonly used method to measure the cosmic star formation history involves determining the rest frame far-ultraviolet luminosity function and correcting it for attenuation (see e.g. \citealt{kennicutt98,madau14}), usually using either the \citet{calzetti00} law or the IRX $(= L_\mathrm{ir}/L_\mathrm{FUV}) - \beta$ relation \citep{meurer99}. For the latter, $A_\mathrm{FUV}$ can be computed directly from IRX. Here we simply compute $A_\mathrm{FUV}$ by convolving the IPEF with the \textit{GALEX} FUV filter curve and converting to magnitudes. Figure \ref{fig:fuvattenuation} shows predictions of $A_\mathrm{FUV}$ compared to the empirical data from \citet{cucciati12,burgarella13} and \citet{andrews16b}. The phenomological model obtains predictions of $A_\mathrm{FUV}$ that are broadly consistent with the empirical data at most redshifts, while \textsc{galform} generally underpredicts $A_\mathrm{FUV}$. At $z \sim 4$, the phenomological model appears to predict a lower $A_\mathrm{FUV}$ than the literature. However, at these redshifts, dust attenuation starts to deviate from the IRX-$\beta$ relation \citep{capak15}. These redshifts are beyond the scope of this paper.

\subsection{The integrated galactic light}
\label{sec:ebl}

\begin{figure*}
\begin{minipage}{7in}
\begin{center}
\includegraphics[width=0.99\linewidth]{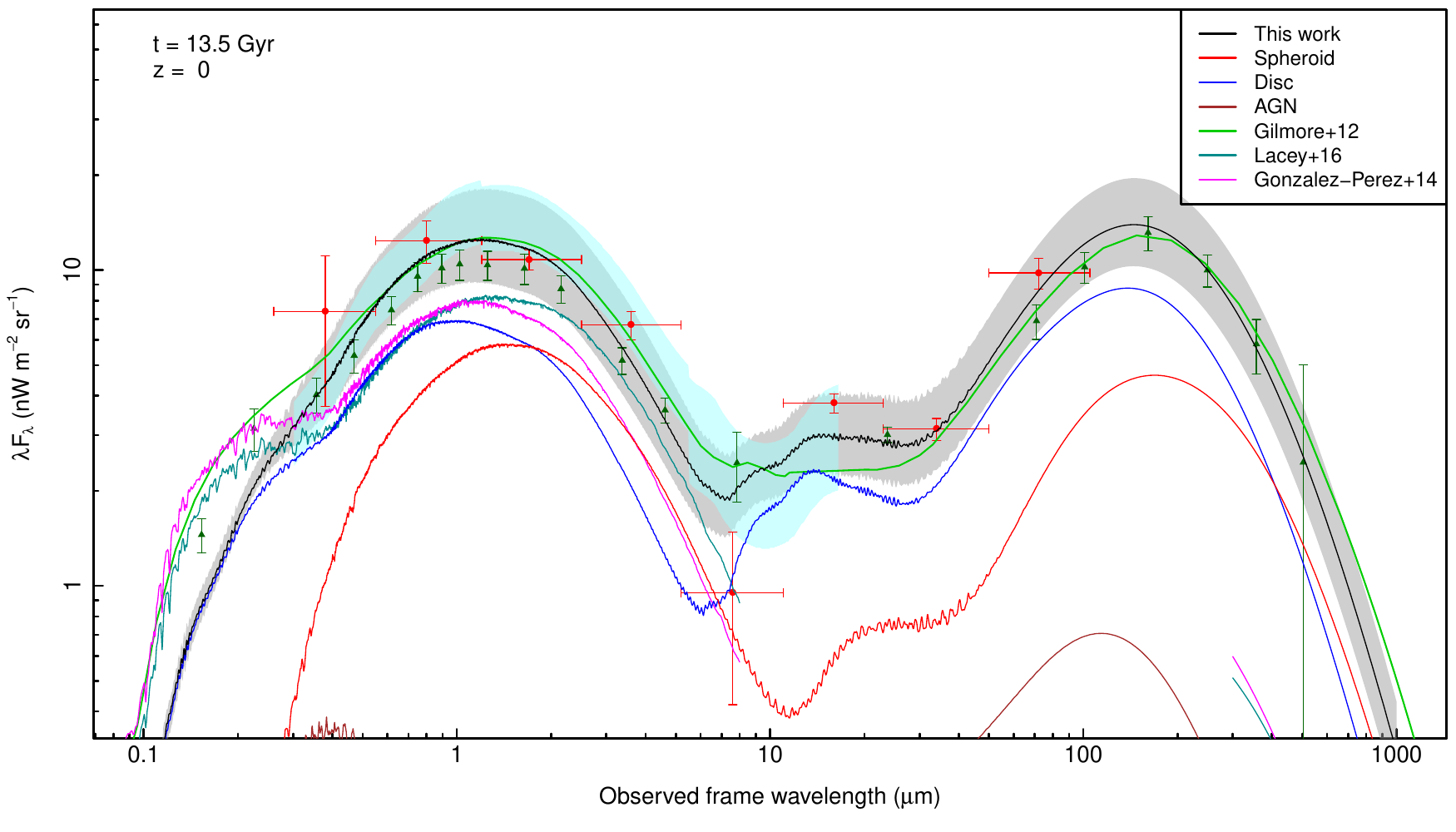}
\caption{The IGL and EBL at $z=0$. Black curve and band: total IGL produced by the model in Section \ref{sec:toymodel} with uncertainty, blue: IGL attributed to discs, red: spheroids, brown: AGN, green: the \citet{gilmore12} semi-analytic model, pink: the \citet{gonzalezperez14} semi-analytic model and orange: the \citet{lacey16} semi-analytic model. The \citet{driver16b} measurements are the dark green triangles, with measurements of the EBL from of $\gamma$-ray data by \citet{biteau15} the red points. The cyan region represents the $1\sigma$ constraints from \citet{hess13}.}
\label{fig:ebl}
\end{center}
\end{minipage}
\end{figure*}

We now look to compare the predicted IGL from our model, as well as from the semi-analytic models to the recent IGL analysis of \citet{driver13}. We also compare against the \citet{gilmore12} predictions from the semi-analytic model of \citet{somerville08} and \citet{somerville12}. In short, this model uses a \citet{chabrier03} IMF, the \citet{bc03} stellar libraries, a modified \citet{charlot00} dust attenuation model and far-infrared spectra from \citet{rieke09}, which are derived from (ultra)-luminous infrared and low-redshift galaxies. 

The comoving EBL at redshift $z$ is the amount of radiation received by an observer at a given epoch and may be derived from the (total) CSED as follows:

\begin{equation}
\mathrm{EBL}(\lambda_\mathrm{obs}, z) = \sum_{z'=z}^{z'=\infty} \frac{\epsilon(\lambda(1+z'), z') dV(z')}{4 \pi d_l(z')^2 }
\end{equation}

where $\epsilon(\lambda, z$) is the rest frame attenuated CSED at the redshift $z$, $d_l(z')$ is the luminosity distance at $z$ to the volume corresponding to the redshift $z'$, and $dV(z')$ is the differential comoving volume of each model timestep subtending a solid angle of 1~sr. 

Figure \ref{fig:ebl} shows a comparison of our phenomological model from Section \ref{sec:toymodel} to measurements of the IGL at $z=0$ \citep{driver16b}. Note that the phenomological model curves are bumpy at wavelengths corresponding to the redshifted Lyman break and emission from polycyclic aromatic hydrocarbons due to the EBL summation involving discrete timesteps. The light blue shaded region denotes constraints on the EBL from the H.E.S.S. \citep{hess13} TeV gamma ray observatory. These observations are able to constrain the amplitude of the cosmic optical background, but draw upon the \citet{franceschini08} and \citet{dominguez11} models respectively to define the shape. The red points represent $\gamma$-ray observations from \citet{biteau15}, which do not assume an EBL model spectrum.

As expected, emission from low-redshift discs dominates at ultraviolet, optical and far-infrared wavelengths. Longwards of $350~\mu$m, we find that high-redshift spheroids represent the dominating contribution to the far-infrared IGL. In the near-infrared, spheroids at low and high redshift are the dominating contribution to the IGL. AGN make a small, but noticable contribution of 0.34~nW~m$^{-2}$~sr$^{-1}$ in the $u$ band, and 0.7~nW~m$^{-2}$~sr$^{-1}$ at $100~\mu$m.

The phenomological model is consistent with the \citet{driver16b} IGL measurements and the \citet{gilmore12} semi-analytic model except in the ultraviolet. The phenomological model becomes consistent with \citet{hess13} at $\lambda > 300$~nm, suggesting an origin at low to intermediate redshifts.

\begin{figure*}
\begin{minipage}{7in}
\begin{center}
\includegraphics[width=0.99\linewidth]{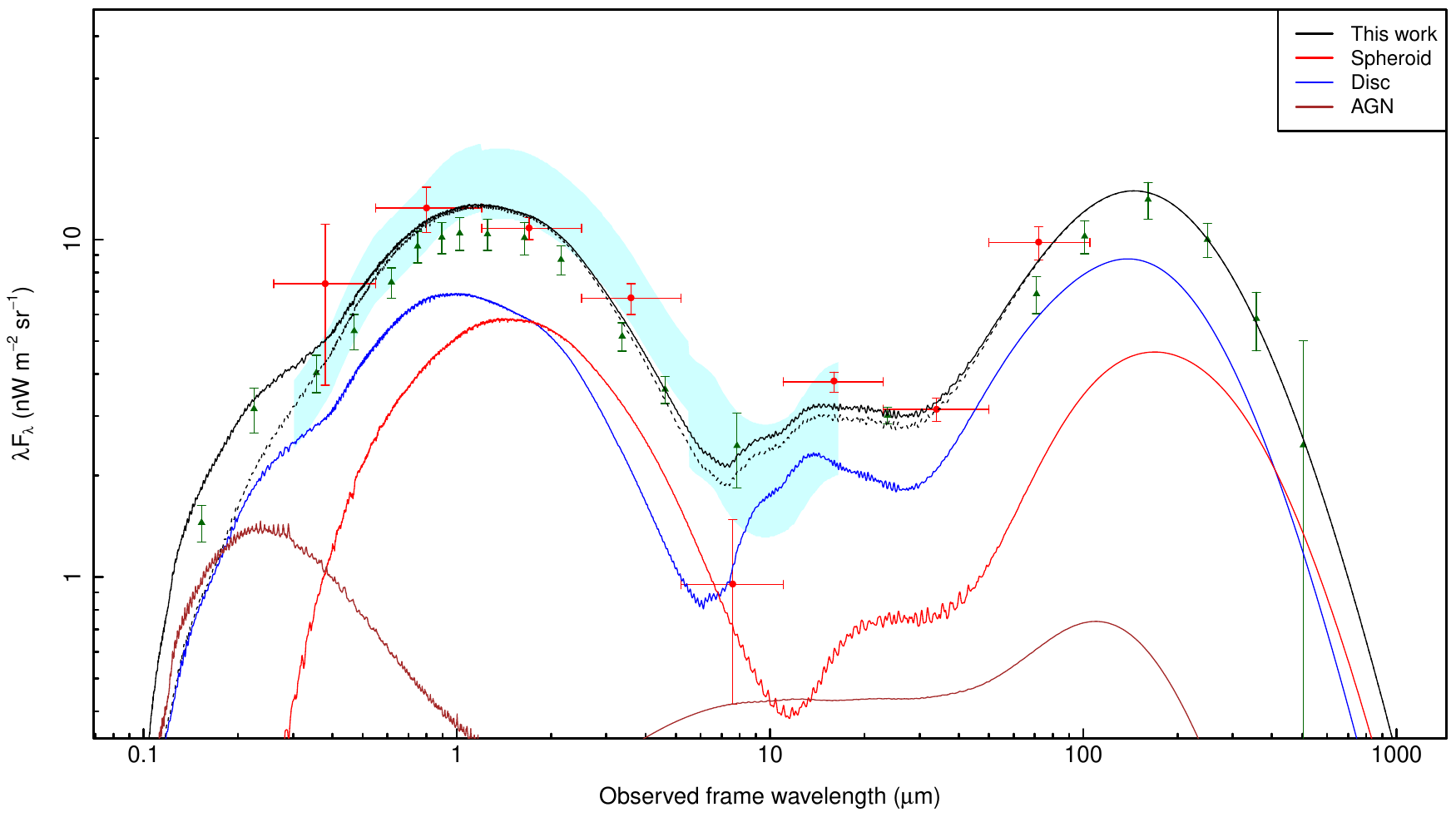}
\caption{The IGL and EBL at $z=0$, as in Figure \ref{fig:ebl}, with an additional component from low-level AGN. The solid black line represents the new model curve, the dashed line the model curve in Figure \ref{fig:ebl} for comparison and brown is the enhanced AGN component.}
\label{fig:ebl2}
\end{center}
\end{minipage}
\end{figure*}

However, there exists some scope for an upward adjustment in the (unobscured) AGN contribution to the EBL, given the \citet{palanque16} measurements of the quasar luminosity density stop at $z = 0.68$ and are associated with an uncertainty of a factor of two at these redshifts. Another potential upward adjustment arises from low-level AGN activity at low and intermediate redshifts. The phenomological model links high-level AGN activity to star formation, resulting in the steep drop off with redshift (see Figure \ref{fig:csfh}). The AGN radio luminosity density for low-luminosity ($L < 10^{25}$~W~Hz$^{-1}$) sources declines relatively slowly with redshift -- $L \propto (1+z)^{1-2.5}$ \citep{smolcic09,mcalpine13,padovani15} -- a rate slower than, or similar to the decline in the total cosmic star formation history. For illustrative purposes, adding a low-power unobscured AGN component with a $g$ band luminosity of $\sim 9 \times 10^{32}$~W~Mpc$^{-3}$ at $z = 0$ (i.e. $\sim3$ per cent of the convolved, attenuated CSED in $g$ at $z = 0.05$) evolving as $(1 + z)^{1.5}$ to $z = 1.5$ (with no contribution prior to $z=1.5$) is sufficient to reproduce the characteristic shape of the ultraviolet IGL. Figure \ref{fig:ebl2} shows this spectrum and demonstrates the potential role of low-level AGN activity in keeping the low redshift Universe ionised. Further work on quantifying their contribution to the low-redshift CSED is clearly warranted.

We estimate the potential upward adjustments from incompleteness and AGN to be 0--2~nW~m$^{-2}$~sr$^{-1}$.  Despite this, there still exists significant tension between the IGL prediction from the phenomological model and the direct \citet{mattila17} EBL estimate using dark cloud observations. A diffuse component to the EBL originating at low to intermediate redshifts cannot be excluded.

In the optical, the phenomological model is consistent with the $\gamma$-ray measurements of the EBL while exceeding the \citet{driver16b} IGL measurements. This extra light could potentially originate from stripped older stellar populations in the halo environment, however a more likely explanation is SED modelling and other forms of error inherent in the phenomological model. In the near-infrared, the phenomological model falls $\sim1\sigma$ below the H.E.S.S. measurements. This could be tentative evidence for diffuse emission from the epoch of reionisation. Most likely, this simply reflects a discrepency between the adopted EBL model used by the $\gamma$-ray measurements. Certainly it would be very useful to recompute the H.E.S.S. and MAGIC constraints using our model.

We also compute the IGL from the mock \textsc{galform} photometry. Here we elect to show the raw predictions, as computed from the discrete CSEDs shown in Figure \ref{fig:mfcsed}. These are interpolated using a 24 point spline and summed over redshift accordingly. Where a CSED is not predicted or interpolated, i.e. shortwards of the pivot wavelength of the \textit{GALEX} FUV filter (153.5~nm) and longwards of $870~\mu$m, it is set to zero. As a result, the model curves show a characteristic sawtooth shape at short wavelengths due to summation over discrete timesteps.

\textsc{galform} is able to predict the far-infrared CSED longward of 70~$\mu$m, but the corresponding predictions of the cosmic infrared background and IGL depend on how the non-prediction of emission from polycyclic aromatic hydrocarbons is treated. Here we treat them as is -- therefore, performing the IGL summation will result in the predicted cosmic infrared background between $8~\mu$m and $\sim 400~\mu$m being systematically lower than expected as the loss of flux is redshifted as shown in Figure \ref{fig:ebl}. 

Unsurprisingly, both iterations of \textsc{galform} produce a cosmic optical background lower than the \citet{driver16b} IGL measurements. This is a consequence of the underprediction of the cosmic star formation history for $z < 3$. When adjusted for the normalisation of the CSED, the \citet{gonzalezperez14} iteration of \textsc{galform} produces predictions of the optical IGL consistent with the phenomological model and the H.E.S.S. and MAGIC TeV $\gamma$-ray EBL observations.

The \citet{lacey16} semi-analytic model produces significantly more radiation in the near-infrared compared to \citet{gonzalezperez14}. This originates at $z > 1$ -- observe the difference between the \citet{gonzalezperez14} and \citet{lacey16} predictions in Figure \ref{fig:mfcsed}. This has interesting implications for SED modelling at high redshifts, given the use of the \citet{maraston05} stellar libraries in the Lacey model versus the \citet{bc03} libraries for all other models. On that basis, it is too early to pinpoint the exact cause of the discrepancy between the IGL and EBL in the near-infrared -- one would need the total errors in both a high-redshift CSED measurement (from the combination of SED modelling, incompleteness and cosmic sample variance) and the $\gamma$-ray EBL measurements to be less than 5 per cent at any one wavelength. In the future, increasing amounts of TeV data will lead to stronger constraints on the shape of the cosmic optical background -- see \citet{biteau15} for an early compilation of measurements. Additionally, observations of PeV gamma rays will lead to similar constraints on the cosmic infrared background via the same method.

\subsection{The cosmic optical and infrared backgrounds}

\begin{figure*}
\begin{minipage}{7in}
\begin{center}
\includegraphics[width=0.99\linewidth]{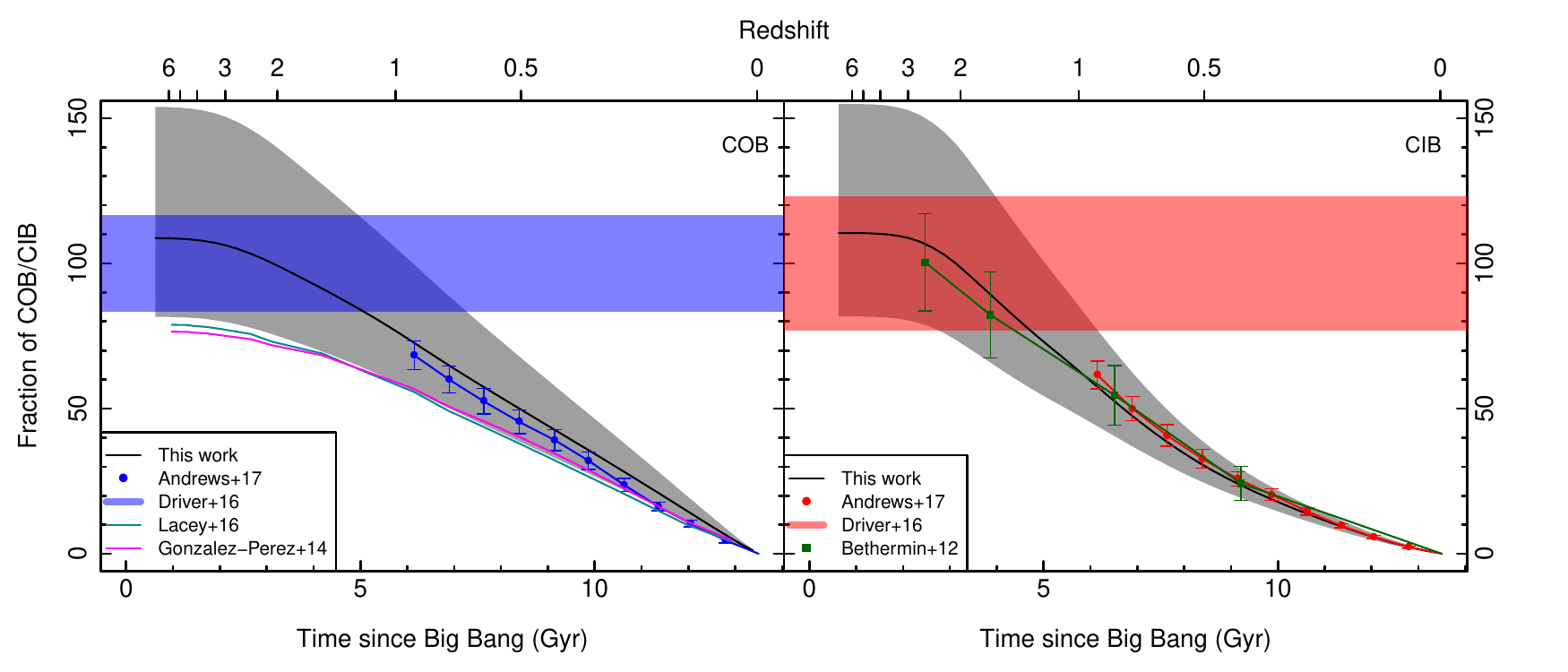}
\caption{Contributions to the COB (left) and CIB (right) as a function of the age of the Universe relative to the total COB and CIB measured by \citet{driver16b}, compared to measurements from \citet{andrews16b} and \citet{bethermin12} and predictions of the COB from \citet{gonzalezperez14} and \citet{lacey16}. Bands indicate the respective uncertainty ranges.}
\label{fig:cobbuildup}
\end{center}
\end{minipage}
\end{figure*}

\begin{figure}
\begin{center}
\includegraphics[width=0.99\linewidth]{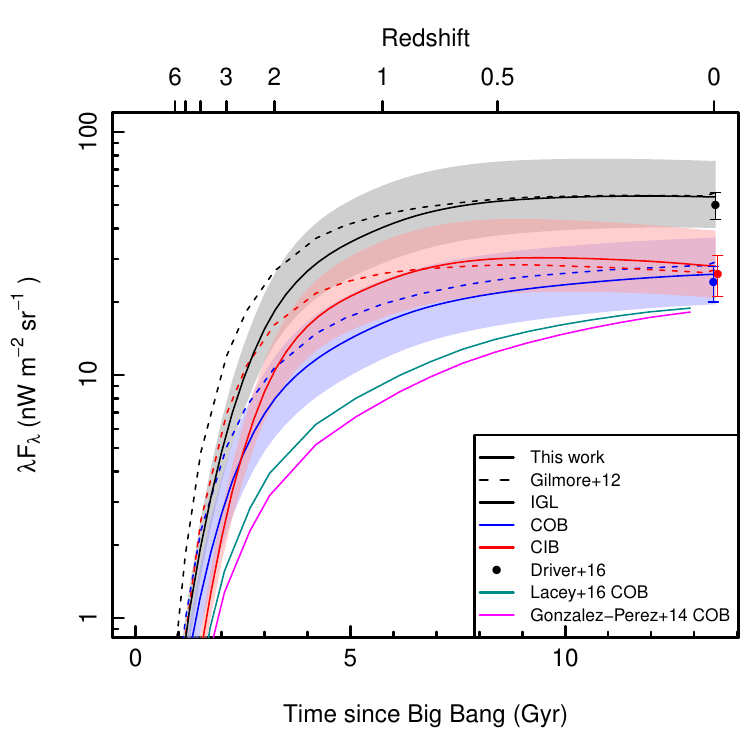}
\caption{The comoving IGL, COB and CIB as a function of cosmic time for the pheomological model and \citet{gilmore12} with predictions for the COB from \citet{gonzalezperez14} and \citet{lacey16}. The \citet{driver16b} measurements at $z = 0$ are shown for comparison. Bands indicate the respective model uncertainty ranges.}
\label{fig:cobevolution}
\end{center}
\end{figure}

We can now break down the EBL into its two key components -- the cosmic optical and infrared backgrounds (COB and CIB respectively) -- by integrating under the respective model curves over the relevant wavelength ranges. Given that \textsc{galform} does not model the mid-infrared, we limit ourselves to comparing the semi-analytic model predictions with the observations in the ultraviolet to the near-infrared only.

Our model (see Figure \ref{fig:cobbuildup}) predicts a roughly equal split between the cosmic optical and infrared backgrounds, in line with observations (see \citealt{driver16b} and compilation within). We predict integrated values for the COB and CIB of $26.0_{-6.5}^{+10.7}$ and $28.0_{-7.2}^{+11.2}$~nW~m$^{-2}$~sr$^{-1}$ respectively for the current epoch, and a peak value for the COB of $12.6_{-3.4}^{+5.6}$~nW~m$^{-2}$~sr$^{-1}$ at 1.19~$\mu$m at the current epoch. This is in agreement with H.E.S.S. and MAGIC, which find maximum values of $15.0 \pm 3.6$ and $12.75^{+2.75}_{-2.29}$~nW~m$^{-2}$~sr$^{-1}$ respectively, both at 1.4~$\mu$m. The \citet{gonzalezperez14} iteration of \textsc{galform} underpredicts the integrated COB by about 35 per cent, yielding an integrated value of 15.8~nW~m$^{-2}$~sr$^{-1}$, while the \citet{lacey16} model predicts 16.7~nW~m$^{-2}$~sr$^{-1}$. Both models also reach peak values of the COB below the H.E.S.S. and MAGIC observations -- 7.7~nW~m$^{-2}$~sr$^{-1}$ at $\lambda = 1.18~\mu$m and 8.0~nW~m$^{-2}$~sr$^{-1}$ at $1.36~\mu$m for the Gonzalez-Perez and Lacey models respectively. For comparison, the \citet{gilmore12} model obtains 29.0 and 27.0~nW~m$^{-2}$~sr$^{-1}$ for the integrated COB and CIB respectively, and predicts a peak value of the COB of 12.7~nW~m$^{-2}$~sr$^{-1}$ at 1.28~$\mu$m. Our model predicts an AGN contribution of approximately 0.6~nW~m$^{-2}$~sr$^{-1}$ or 2.5 per cent (approximately equally split between obscured and unobscured AGN) to the integrated COB and 1.4~nW~m$^{-2}$~sr$^{-1}$ or 5.1 per cent to the integrated cosmic infrared background (mostly from obscured AGN).

Finally, our model predicts a CIB of $0.53_{-0.15}^{+0.24}$~nW~m$^{-2}$~sr$^{-1}$ at $850~\mu$m, in line with the $0.43_{-0.15}^{+0.24}$~nW~m$^{-2}$~sr$^{-1}$ observed using galaxy number counts from SCUBA-II \citep{chen13b} and $0.50_{-0.19}^{+0.23}$~nW~m$^{-2}$~sr$^{-1}$ from direct observations using the \textit{Cosmic Background Explorer} \citep{fixsen98}. The \citet{gonzalezperez14} and \citet{lacey16} models also show good agreement, predicting 0.44 and 0.54~nW~m$^{-2}$~sr$^{-1}$ respectively. The CIB at this wavelength primarily originates from galaxies at $z = 1-3$ \citep{zavala17}, lending confidence to the predictions of all models at higher redshifts.

Figure \ref{fig:cobbuildup} shows the accumulation of the COB and the CIB as one looks back in time. All model contributions to the CIB agree very well with both the \citet{andrews16b} and \citet{bethermin12} measurements, while the phenomological model slightly overpredicts the COB relative to \citet{andrews16b}. The phenomological model reaches final values of the COB and CIB fully consistent with \citet{driver16b}. The Figure also shows the underprediction of the COB by \textsc{galform}. This is not surprising given the deficit of the \textsc{galform} predicted CSED below the phenomological model (see Figures \ref{fig:mucsed} and \ref{fig:mfcsed}), supporting the predictions of the phenomological model. 

Figure \ref{fig:cobevolution} shows the advantage of CSED measurements at multiple epochs over comoving IGL measurements at $z=0$ for constraining galaxy evolution models -- the CSED measurements are able to constrain the gradient of this curve at multiple epochs, whereas the COB, CIB, IGL and EBL at these epochs is unobservable. Both our model and \citet{gilmore12} give similar values for the present-day IGL, COB and CIB but differ at higher redshifts. We find that the comoving COB increases monotonically with time, while the CIB peaks about 9.25 Gyr, diminishing very slowly thereafter. The total IGL initially increases rapidly, before levelling off at $\sim51$~nW~m$^{-2}$~sr$^{-1}$ by approximately 9~Gyr. The \citet{gilmore12} model obtains similar predictions, with the CIB peaking at a slightly earlier time. Like all cosmic background radiation, the integrated IGL, EBL, COB and CIB will diminish over time due to redshifting if it is not maintained by further energy production pathways.

\subsection{Stellar mass growth}
\label{sec:stellarmass}

\begin{figure}
\begin{center}
\includegraphics[width=0.99\linewidth]{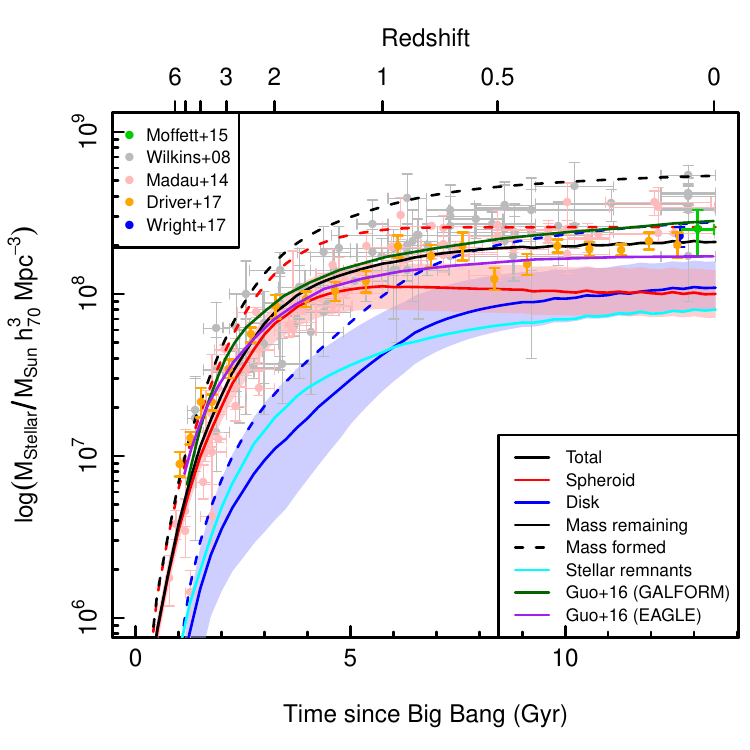}
\caption{Stellar mass as a function of cosmic time in the empirical model, with mass formed (dashed lines) and mass surviving (solid lines, with bands to indicate the model uncertainty) for spheroids (red), discs (blue) and both combined (black). The total mass in stellar remnants for both components is shown in cyan. The green, blue and orange points represent the \citet{moffett16}, \citet{wright17} and \citet{driver17} stellar mass measurements respectively, while the faint red and grey points represent the \citet{madau14} and \citet{wilkins08} compilations. Also shown are predictions from \textsc{galform} (dark green) and \textsc{eagle} (purple).}
\label{fig:stellarmass}
\end{center}
\end{figure}

While the phenomological model is focused and calibrated to energy, it is also capable of providing a complete description of stellar and dust mass evolution.

Figure \ref{fig:stellarmass} shows the predicted buildup of stellar mass as a function of time for spheroids and discs (more precisely, the material that exists in spheroids and discs today). We show two curves -- total stellar mass formed, computed by integrating the CSFH fitting functions (dashed lines, with bands to indicate the model uncertainty) and mass surviving at the specified time (solid lines). The curves differ as stellar material is returned to the ISM by supernovae and winds of thermally-pulsating asymptotic giant branch stars. The \citet{bc03} models return from 35 per cent at 1~Gyr to 44 per cent at 10~Gyr of stellar mass to the ISM and place 8 per cent at 1~Gyr to 14 per cent at 10~Gyr into stellar remnants.

The green and blue points represents the total stellar mass density as reported by \citet{moffett16} and \citet{wright17}. The \citet{moffett16} integrated stellar mass estimates have a 32 per cent uncertainty, equally portioned (22 per cent each) between uncertainty in the fitted functions and cosmic sample variance. The model predicts an almost equal distribution of stellar mass residing in discs (49.9 per cent) and spheroids (50.1 per cent), reproducing the 47 to 53 per cent stellar mass ratio of \citet{moffett16b}. (`Little blue spheroids' are deemed to be disc material.) We do not have to adjust the spheroid star formation history by 25 per cent akin to \citet{driver13}. Note that the spheroid to disc mass ratio is independent of cosmic variance, barring large hidden clustering effects. We also note that the total stellar mass in spheroids enters a slow decline since $t=5$~Gyr with the total stellar mass levelling off at $t \sim$10~Gyr, in agreement with \citet{driver13}.

We supplement Figure \ref{fig:stellarmass} with predictions derived by \citet{guo16} from the \citet{gonzalezperez14} version of \textsc{galform} and the \textsc{eagle} suite of hydrodynamic simulations \citep{schaye15}. Briefly, the \textsc{eagle} simulation used by \citet{guo16} consists of a (100 cMpc)$^3$ box populated by 1504$^3$ particles of both gas and dark matter. \textsc{eagle} uses a \citet{chabrier03} IMF and the \citet{bc03} stellar libraries to derive mock photometry and stellar masses. The phenomological model predicts stellar mass equally as good as \textsc{galform}, and matches the low-redshift data better than \textsc{eagle}.

\subsection{Dust mass growth}
\label{sec:dustmass}

\begin{figure}
\begin{center}
\includegraphics[width=0.99\linewidth]{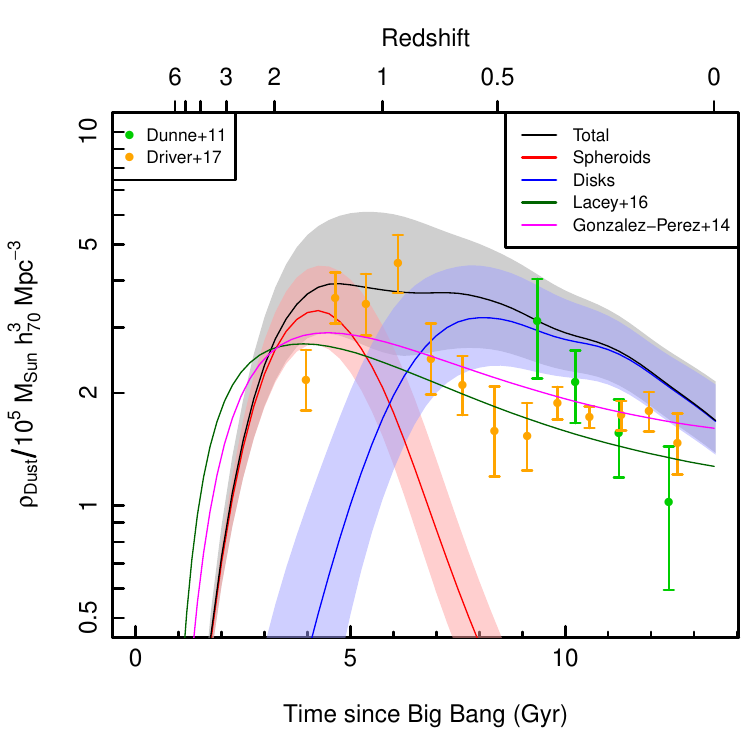}
\caption{Dust mass density as a function of cosmic time in the empirical model, with dust mass in spheroids, discs and both combined (red, blue and black respectively, with bands to indicate uncertainty), with observations by \citet{dunne11} (green) and \citet{driver17} (orange) and predictions from \textsc{galform} -- \citet{gonzalezperez14} in pink and \citet{lacey16} in dark green.}
\label{fig:dustmass}
\end{center}
\end{figure}

Using the phenomological model, we can also arrive at predictions of the dust mass density from the normalisation constant used to scale up the \citet{dale14} templates to the total energy absorbed and re-released into the far-infrared. Two parameters are necessary to do so -- the mass of a gas particle with respect to the mass of a hydrogen atom $m_\mathrm{gas}/m_H$ and the dust to gas mass ratio $M_\mathrm{dust}/M_\mathrm{gas}$. We assume $m_\mathrm{gas}/m_H = 1.3$ and $M_\mathrm{dust}/M_\mathrm{gas}$ that scales linearly with the generating function for the metallicity of newly formed stars (i.e. the total stellar mass formed, see Figure \ref{fig:metallicity}) to reach $M_\mathrm{dust}/M_\mathrm{gas} = 10^{-2.5}$ at $z = 0$. The assumed $M_\mathrm{dust}/M_\mathrm{gas}$ ratio at $z = 0$ value is within the range of $M_\mathrm{dust}/M_\mathrm{gas}$ observed in nearby galaxies by the \textit{Herschel} Reference Survey \citep{cortese16}, while $m_\mathrm{gas}/m_H = 1.3$ is assumed by the same to account for helium in the ISM.

Figure \ref{fig:dustmass} shows these predictions compared to the \citet{driver17} and \citet{dunne11} observations, as well as \textsc{galform}. We smooth our model predictions with a ten-point spline to remove discretisation caused by the abrupt shifts in the choice of \citet{dale14} template spectra. For $0.45 < z < 1.75$, the fraction of the total dust mass density resulting from predictions and low-significance far-infrared observations increases with redshift (see e.g. Figure 7 of \citealt{andrews16b}). The phenomological model predicts higher dust mass densities at low redshifts than literature measurements and \textsc{galform}. Given the large uncertainties in the \citet{andrews16b} far-infrared CSEDs, the fraction of the total dust mass density arising from predictions, the underlying shallowness of the far-infrared photometry (especially at $70~\mu$m rest frame), and the lack of a comprehensive collection of literature measurements of the dust mass density (akin to the cosmic star formation history), this overprediction is not a concern. 

The phenomological model predicts that elliptical galaxies become mostly devoid of dust by 8~Gyr. This is consistent with the 76 per cent non-detection rate of elliptical galaxies in the \textit{Herschel} Reference Survey \citep{smith12} and 94.5 per cent in H-ATLAS \citep{rowlands12}. The small number of elliptical galaxies with detectable dust can be plausibly explained by the result of accretion from mergers with star-forming galaxies, an effect not accounted for in the phenomological model and tentatively suggested by the observations.

\section{Conclusion}
\label{sec:conclusion}

We have extended the \citet{driver13} model to produce predictions of the CSED and IGL from the far-ultraviolet to the far-infrared over the life of the Universe to incorporate dust attenuation and emission from AGN. The model is based on simple expressions of the cosmic star formation history as attributed to spheroids and discs, with the speheroid star formation linked to AGN activity and dominating at high-redshift. We use the \citet{bc03} stellar libraries, the \citet{charlot00} dust attenuation prescription and the \citet{dale14} dust emission spectra in order to reproduce the \citet{andrews16b} CSED estimates at low and intermediate redshifts. We then link spheroid star formation to AGN activity and derive spectra for unobscured and obscured AGN using the \citet{vandenberk01} composite spectrum to derive predictions of the CSED at high redshifts. This model is able to reproduce observations of the IGL, stellar and dust mass densities and the fractions of stellar mass in bulges and discs.

We have also obtained equivalent semi-analytic predictions of the CSED using \textsc{galform} \citep{gonzalezperez14,lacey16}. We find that \textsc{galform} is able to reproduce the shape of both the unattenuated and attenuated CSEDs well over $0 < z < 1$, but underpredicts the normalisation due to underpredicting the cosmic star formation history \citep{mitchell14,guo16,lacey16}. \textsc{galform} does not model the mid-infrared emission due to the lack of a prescription for emission and absorption from polycyclic aromatic hydrocarbons and underestimates the amount of dust attenuation in the ultraviolet, presumably due to galaxy half-mass radii being overpredicted on average \citep{merson16}. 

Our model does not include a clustering prescription by design. We will examine the CSED as a function of environment using the GAMA group \citep{robotham11} and large scale structure \citep{alpaslan14} catalogues and their intermediate redshift equivalents in a future paper. Other considerations, such as the CSED as a function of morphology and inclination, will also be examined in future works.

The attribution of the cosmic star formation history to spheroids and discs may be verified using accurate bulge-disc decompositions of both samples where permitted by image resolution. While such a sample currently exists in GAMA at low redshift ($z < 0.06$), there are still reasonable doubts as to whether bulge-disc decomposition codes are able to converge on a physically meaningful model given an arbitrary or empty set of initial conditions \citep{lange16}. We are currently extending this sample out to $z < 0.1$ using data from the Kilo-Degree Survey \citep{dejong15} and an improved bulge-disc decomposition algorithm \citep{robotham17}. Finely gridded stellar population models will increase the precision of modelling stellar phase metallicity, with the caveat that additional discretisation and interpolation effects may arise from coarseness in the underlying isochrones. Deeper far-infrared data, especially $70~\mu$m, is required to derive firm measurements on the far-infrared CSED, rigourously optimise fitting of template spectra to the measurements and derive corresponding predictions of dust temperature distributions. The incorporation of AGN templates into \textsc{magphys}, a reduction of SED modelling error, cosmic sample variance and incompleteness by a factor of about four each precise measurements of the CSED at higher redshifts and are all required to derive strong constraints on the contribution of AGN to the CSED, the contribution of diffuse emission to the (total) CSED and EBL and the shape and normalisation of the near-infrared IGL and EBL.

In the future, radio data from the Australian Square Kilometre Array Pathfinder, \textit{Spektr-RG}'s eROSITA and the COSMOS HI Large Extragalactic Survey will become available for the GAMA and G10/COSMOS regions. This data, in combination with future improvements to SED fitting tools to accommodate the expanded wavelength range, will increase our understanding of the gas content and AGN activity and hence allow us to extend our models and measurements of the CSED and IGL to radio wavelengths. In addition, the Wide Area VISTA Extragalactic Survey \citep{waves} will reduce the errors associated with cosmic sample variance and incompleteness in the empirical CSED measurements. Data from the Cherenkov Telescope Array will help reduce the systematic error inherent in the $\gamma$-ray EBL measurements. Finally, the \textit{James Webb Space Telescope}, \textit{Wide-Field Infrared Space Telescope} and \textit{Euclid} will allow investigations at higher redshift, thus fully constraining the optical and near-infrared CSED across the entire history of the Universe.

\section*{Acknowledgements}

We thank the referee for an insightful report that helped to improve this manuscript. SKA is supported by the Australian Government’s Department of Industry Australian Postgraduate Award (APA) and a travel grant from the Convocation of UWA Graduates. He also wishes to thank the University of St. Andrews, where the majority of this work was done, for their warm hospitality.

\section*{Supporting information}
Animated versions of Figures \ref{fig:mucsed} (mucsed.gif) and \ref{fig:mfcsed} (mfcsed.gif) are available online as supporting information. These animations also include a comparison to the \citet{driver12} CSED measurements (discs: orange points, spheroids: brown points).
\label{lastpage}


\begin{thebibliography}{99}
\bibitem[\protect\citeauthoryear{Ahnen et al.}{2016}]{magic16} Ahnen M.~L. et al. 2016, A\&A, 590, A24
\bibitem[\protect\citeauthoryear{Allard \& Hauschildt}{1995}]{allard95} Allard F. \& Hauschildt P.~H. 1995, ApH, 445, 433
\bibitem[\protect\citeauthoryear{Alpaslan et al.}{2014}]{alpaslan14} Alpaslan M. et al. 2014, MNRAS, 438, 177
\bibitem[\protect\citeauthoryear{Andrews et al.}{2017a}]{andrews16a} Andrews S.~K., Driver S.~P., Davies L.~J~.M., Kafle P.~R., Robotham A.~S.~G., Wright A.~H. 2017a, MNRAS, 464, 1569
\bibitem[\protect\citeauthoryear{Andrews et al.}{2017b}]{andrews16b} Andrews S.~K. et al. 2017b, MNRAS, 470, 1342
\bibitem[\protect\citeauthoryear{Arendt \& Dwek}{2003}]{arendt03} Arendt R. G., \& Dwek E. 2003, ApJ, 585, 305
\bibitem[\protect\citeauthoryear{Ashby et al.}{2013}]{ashby13} Ashby M. L. N. et al. 2013, ApJ, 769, 80
\bibitem[\protect\citeauthoryear{Baldry \& Glazebrook}{2003}]{baldry03} Baldry I. K. \& Glazebrook K., 2003, ApJ, 593, 258
\bibitem[\protect\citeauthoryear{Bernstein et al.}{2002}]{bernstein02} Bernstein R., Freedman W.~L., Madore B.~F. 2002, ApJ, 571, 56
\bibitem[\protect\citeauthoryear{Bernstein}{2007}]{bernstein07} Bernstein, R. A. 2007, ApJ, 666, 663
\bibitem[\protect\citeauthoryear{Berta et al.}{2011}]{berta11} Berta S. et al. 2011, A\&A 532, A49
\bibitem[\protect\citeauthoryear{B\'{e}thermin et al.}{2010}]{bethermin10} B\'{e}thermin M., Dole H., Beelan A., Aussel H. 2010, A\&A, 512, A78
\bibitem[\protect\citeauthoryear{B\'{e}thermin et al.}{2012a}]{bethermin12} B\'{e}thermin M., et al. 2012a, A\&A, 542, A58
\bibitem[\protect\citeauthoryear{B\'{e}thermin et al.}{2012b}]{bethermin12b} B\'{e}thermin M. et al. 2012b, ApJ, 757, L23
\bibitem[\protect\citeauthoryear{B\'{e}thermin et al.}{2014}]{bethermin14} B\'{e}thermin M. et al. 2014, A\&A, 567, A103
\bibitem[\protect\citeauthoryear{Biteau \& Williams}{2015}]{biteau15} Biteau J. \& Williams D.~A. 2015, ApJ, 812, 60
\bibitem[\protect\citeauthoryear{Bourne et al.}{2017}]{bourne17} Bourne N. et al. 2017, MNRAS, 467, 1360
\bibitem[\protect\citeauthoryear{Bruzual}{2007}]{bruzual07} Bruzual, G. 2007, in ASP Conf. Ser. 374, From Stars to Galaxies: Building the Pieces to Build Up the Universe, ed. A. Vallenari et al. (San Francisco, CA: ASP), 303
\bibitem[\protect\citeauthoryear{Bruzual \& Charlot}{2003}]{bc03} Bruzual G. \& Charlot S. 2003, MNRAS, 344, 1000
\bibitem[\protect\citeauthoryear{Bruzual et al.}{2013}]{bruzual13} Bruzual, G., Charlot, S., Gonz\'{a}lez L\'{o}pezlira, R., Srinivasan S., Boyer M.~L., Riebel D. 2013, in Proceedings of the IAU Symposium No. 295 ``The LF of TP-AGB stars in the LMC/SMC" eds. D. Thomas, A. Pasquali \& I. Ferreras, Cambridge: Cambridge University Press, 282
\bibitem[\protect\citeauthoryear{Buchner et al.}{2015}]{buchner15} Buchner J. et al. 2015, ApJ, 802, 89
\bibitem[\protect\citeauthoryear{Burgarella et al.}{2013}]{burgarella13} Burgarella D. et al. 2013, A\&A, 554, A70
\bibitem[\protect\citeauthoryear{Calzetti et al.}{2000}]{calzetti00} Calzetti D., Armus L., Bohlin R.~C., Kinney A.~L., Koornneef J., Storchi-Bergmann T. 2000, ApJ, 533, 682
\bibitem[\protect\citeauthoryear{Capak et al.}{2015}]{capak15}  Capak P. et al. 2015, Nature, 522, 455
\bibitem[\protect\citeauthoryear{Capozzi et al.}{2016}]{capozzi16} Capozzi D., Maraston C., Daddi E., Renzini A., Strazzullo V., Gobat R. 2016, MNRAS, 456, 790
\bibitem[\protect\citeauthoryear{Cardelli et al.}{1989}]{cardelli89} Cardelli J. A., Clayton G. C., Mathis J. S. 1989, ApJ, 345, 245
\bibitem[\protect\citeauthoryear{Carniani et al.}{2015}]{carniani15} Carniani S. et al. 2015, A\&A, 584, A78
\bibitem[\protect\citeauthoryear{Chabrier}{2003}]{chabrier03} Chabrier G. 2003, PASP, 115, 763
\bibitem[\protect\citeauthoryear{Charlot \& Fall}{2000}]{charlot00} Charlot S. \& Fall S.~M. 2000, ApJ, 539, 718
\bibitem[\protect\citeauthoryear{Chary \& Elbaz}{2001}]{chary01} Chary R. \& Elbaz D. 2001, ApJ, 556, 562
\bibitem[\protect\citeauthoryear{Chen et al.}{2013}]{chen13} Chen C.-T. J. et al. 2013, ApJ, 773, 3
\bibitem[\protect\citeauthoryear{Chen et al.}{2013b}]{chen13b} Chen C.-C., Cowie L.~L., Barger A.~J., Casey C.~M., Lee N., Sanders D.~B., Wang W.-H., Williams J.~P., 2013, ApJ, 776, 131
\bibitem[\protect\citeauthoryear{Conroy \& Gunn}{2010}]{conroy10} Conroy C. \& Gunn J.~E. 2010, ApJ, 712, 833
\bibitem[\protect\citeauthoryear{Cortese et al.}{2016}]{cortese16} Cortese L. et al. 2016, MNRAS, 459, 3574
\bibitem[\protect\citeauthoryear{Cowley et al.}{2017}]{cowley17} Cowley W.~I., B\'{e}thermin M., Lagos C.~d.~P., Lacey C.~G., Baugh C.~M., Cole S. 2017, MNRAS, 467, 1231
\bibitem[\protect\citeauthoryear{Cucciati et al.}{2012}]{cucciati12} Cucciati O. et al. 2012, A\&A, 539, A31
\bibitem[\protect\citeauthoryear{da Cunha et al.}{2008}]{dacunha08} da Cunha E., Charlot S., Elbaz D. 2008, MNRAS 388, 1595
\bibitem[\protect\citeauthoryear{Dale \& Helou}{2002}]{dale02} Dale D. \& Helou G., 2002, ApJ, 576, 159
\bibitem[\protect\citeauthoryear{Dale et al.}{2014}]{dale14} Dale D., Helou G., Magdis G.~E., Armus L., D\'{i}az-Santos T., Shi Y. 2014, ApJ, 784, 83
\bibitem[\protect\citeauthoryear{Davies et al.}{2015}]{davies15} Davies L.~J.~M. et al. 2015, MNRAS, 447, 1014
\bibitem[\protect\citeauthoryear{Davies et al.}{2016}]{davies16} Davies L.~J.~M. et al. 2016, MNRAS, 461, 458
\bibitem[\protect\citeauthoryear{de Jong et al.}{2015}]{dejong15} de Jong J. T. A. et al. 2015, A\&A, 582, A62
\bibitem[\protect\citeauthoryear{Dom\'{i}nguez et al.}{2011}]{dominguez11} Dom\'{i}nguez A. et al. 2011, MNRAS, 410, 2556
\bibitem[\protect\citeauthoryear{Draine \& Li}{2007}]{draine07} Draine B.~T. \& Li A. 2007, ApJ, 657, 810
\bibitem[\protect\citeauthoryear{Driver et al.}{2008}]{driver08} Driver S.~P., Popescu C.~C., Tuffs R.~J., Graham A.~W., Liske J., Baldry I.~K. 2008, ApJ, 678, L101
\bibitem[\protect\citeauthoryear{Driver et al.}{2011}]{driver11} Driver S.~P. et al. 2011, MNRAS, 413, 971
\bibitem[\protect\citeauthoryear{Driver et al.}{2012}]{driver12} Driver S.~P. et al. 2012, MNRAS, 427, 3244
\bibitem[\protect\citeauthoryear{Driver et al.}{2013}]{driver13} Driver S.~P., Robotham A.~S.~G., Bland-Hawthorn J., Brown M., Hopkins A., Liske J., Phillipps S., Wilkins S. 2013, MNRAS, 430, 2622
\bibitem[\protect\citeauthoryear{Driver et al.}{2015}]{waves} Driver S.~P., Davies L.~J., Meyer M., Power C., Robotham A.~S.~G., Baldry I.~K., Liske J., Norberg P. 2016, ASSP, 42, 205
\bibitem[\protect\citeauthoryear{Driver et al.}{2016a}]{driver16a} Driver S.~P. et al. 2016a, MNRAS, 455, 3911
\bibitem[\protect\citeauthoryear{Driver et al.}{2016b}]{driver16b} Driver S.~P. et al. 2016b, ApJ, 827, 108
\bibitem[\protect\citeauthoryear{Driver et al.}{2017}]{driver17} Driver S.~P. et al. 2017, MNRAS, in press
\bibitem[\protect\citeauthoryear{Dunne et al.}{2011}]{dunne11} Dunne L. et al. 2011, MNRAS, 417, 1510
\bibitem[\protect\citeauthoryear{Dwek et al.}{1998}]{dwek98} Dwek E., et al. 1998, ApJ, 508, 106
\bibitem[\protect\citeauthoryear{Elvis et al.}{1994}]{elvis94} Elvis M. et al. 1994, ApJS, 95, 1
\bibitem[\protect\citeauthoryear{Erb et al.}{2006}]{erb06} Erb D., Shapley A., Pettini M., Steidel C. C., Reddy N. A., Adelberger K. L., 2006, ApJ, 644, 813
\bibitem[\protect\citeauthoryear{Finkbeiner et al.}{2000}]{finkbeiner00} Finkbeiner D. P., Davis M., \& Schlegel D. J. 2000, ApJ, 544, 81
\bibitem[\protect\citeauthoryear{Finke et al.}{2010}]{finke10} Finke J.~D., Razzaque S., Dermer C.~D. 2010, ApJ, 712, 238
\bibitem[\protect\citeauthoryear{Fixsen et al.}{1998}]{fixsen98} Fixsen D.~J., Dwek E., Mather J.~C., Bennett C.~L., Shafer R.~A., 1998, ApJ, 508, 123
\bibitem[\protect\citeauthoryear{Franceschini et al.}{2008}]{franceschini08} Franceschini A., Rodighiero G., Vaccari M. 2008, A\&A, 487, 837
\bibitem[\protect\citeauthoryear{Gallazzi et al.}{2009}]{gallazzi09} Gallazzi A., Brinchmann J., Charlot S., White S. D. M. 2009, MNRAS, 383, 1439
\bibitem[\protect\citeauthoryear{Gardner et al.}{2000}]{gardner00} Gardner J.~P., Brown T.~M. \& Ferguson H.~C. 2000, ApJ, 542, L79
\bibitem[\protect\citeauthoryear{Gebhardt et al.}{2000}]{gebhardt00} Gebhardt K. et al. 2000, AJ, 119, 1157
\bibitem[\protect\citeauthoryear{Gil de Paz et al.}{2005}]{gildepaz05} Gil de Paz A. et al. 2005, ApJ, 627, L29
\bibitem[\protect\citeauthoryear{Gilmore et al.}{2012}]{gilmore12} Gilmore R.~C., Somerville R.~S., Primack J.~R., Dom\'{i}nguez A. 2012, MNRAS, 422, 3189
\bibitem[\protect\citeauthoryear{Gonz\'{a}lez Delgado et al.}{2015}]{gonzalez15} Gonz\'{a}lez Delgado R.~M. et al. 2015, A\&A, 581, A103
\bibitem[\protect\citeauthoryear{Gonzalez-Perez et al.}{2014}]{gonzalezperez14} Gonzalez-Perez V., Lacey C.~G. Baugh C.~M., Lagos C.~D.~P., Helly J., Campbell D.~J.~R., Mitchell P.~D. 2014, MNRAS, 439, 264
\bibitem[\protect\citeauthoryear{Gonzalez-Perez et al.}{2017}]{gonzalezperez17} Gonzalez-Perez V. et al. 2017, MNRAS, submitted
\bibitem[\protect\citeauthoryear{Gordon et al.}{2003}]{gordon03} Gordon K.~D., Clayton G.~C., Misselt K.~A., Landolt A.~U., Wolff M.~J. 2003, ApJ, 594, 279
\bibitem[\protect\citeauthoryear{Graham \& Scott}{2015}]{graham15} Graham A.~W. \& Scott N. 2015, ApJ, 798, 54
\bibitem[\protect\citeauthoryear{Guo et al.}{2016}]{guo16} Guo Q. et al. 2016, MNRAS, 461, 3457
\bibitem[\protect\citeauthoryear{Hauser et al.}{1998}]{hauser98} Hauser M.~G. et al. 1998, ApJ, 508, 25
\bibitem[\protect\citeauthoryear{H.E.S.S. Collaboration}{2013}]{hess13} H.E.S.S. Collaboration 2015 A\&A, 550, A4
\bibitem[\protect\citeauthoryear{Hopkins \& Beacom}{2006}]{hopkins06} Hopkins A.~M., Beacom J. F. 2006, ApJ, 651, 142
\bibitem[\protect\citeauthoryear{Ibarra-Medel et al.}{2016}]{ibarramedel16} Ibarra-Medel H. J. et al. 2016, MNRAS, 463, 2799
\bibitem[\protect\citeauthoryear{Inoue et al.}{2013}]{inoue13} Inoue Y., Inoue S., Kobayashi M.~A.~R., Makiya R., Niino Y., Totani T. 2013, ApJ, 768, 197
\bibitem[\protect\citeauthoryear{Jauzac et al.}{2011}]{jauzac11} Jauzac M. et al. 2011, A\&A, 525, A52
\bibitem[\protect\citeauthoryear{Kawara et al.}{2017}]{kawara17} Kawara K., Matsuoka Y., Sano K., Brandt T.~D., Sameshima H., Tsumura K., Oyabu S., Ienaka N. 2017, PASJ, in press
\bibitem[\protect\citeauthoryear{Keenan et al.}{2010}]{keenan10} Keenan R.~C., Barger A.~J., Cowie L.~L., Wang W.-H. 2010, ApJ, 723, 40
\bibitem[\protect\citeauthoryear{Kelvin et al.}{2014}]{kelvin14} Kelvin L.~S. et al. 2014, MNRAS, 439, 1245
\bibitem[\protect\citeauthoryear{Kennicutt}{1983}]{kennicutt83} Kennicutt R.~C. 1983, ApJ, 272, 54
\bibitem[\protect\citeauthoryear{Kennicutt}{1998}]{kennicutt98} Kennicutt R.~C. 1998, ApJ, 498, 541
\bibitem[\protect\citeauthoryear{Khaire \& Srianand}{2015}]{khaire15} Khaire V. \& Srianand R. 2015, ApJ, 805, 33
\bibitem[\protect\citeauthoryear{Kormendy \& Ho}{2013}]{kormendy13} Kormendy J. \& Ho L. C. 2013, ARA\&A, 51, 511
\bibitem[\protect\citeauthoryear{Kroupa}{2001}]{kroupa01} Kroupa P. 2001, MNRAS, 322, 231
\bibitem[\protect\citeauthoryear{Lacey et al.}{2016}]{lacey16} Lacey C. et al. 2016, MNRAS, 462, 3854
\bibitem[\protect\citeauthoryear{Lagos et al.}{2014}]{lagos14} Lagos C.~d.~P., Baugh C. M., Zwaan M. A., Lacey C. G., Gonzalez-Perez V., Power C., Swinbank A. M., van Kampen E. 2014, MNRAS, 440, 920
\bibitem[\protect\citeauthoryear{Lange et al.}{2016}]{lange16} Lange R. et al. 2016, MNRAS, 462, 1470
\bibitem[\protect\citeauthoryear{Le Borgne et al.}{2004}]{leborgne04} Le Borgne D., Rocca-Volmerange, B., Prugniel P., Lan\c{c}on A., Fioc M., Soubiran, C. 2004, A\&A, 425, 881
\bibitem[\protect\citeauthoryear{Levenson et al.}{2007}]{levenson07} Levenson L. R., Wright E. L. \& Johnson, B. D. 2007, ApJ, 666, 34
\bibitem[\protect\citeauthoryear{Levenson \& Wright}{2008}]{levenson08} Levenson L.~R., Wright E.~L., 2008, ApJ, 683, 585
\bibitem[\protect\citeauthoryear{Lilly et al.}{1996}]{lilly96} Lilly S.~J., Le Fevre O., Hammer F., Crampton D. 1996, ApJ, 460, L1
\bibitem[\protect\citeauthoryear{Liske et al.}{2015}]{liske15} Liske J. et al. 2015, MNRAS, 452, 2087
\bibitem[\protect\citeauthoryear{Lusso et al.}{2013}]{lusso13} Lusso E. et al. 2013, ApJ, 777, 86
\bibitem[\protect\citeauthoryear{Madau \& Dickinson}{2014}]{madau14} Madau P. \& Dickinson M. 2014, ARA\&A, 52, 415
\bibitem[\protect\citeauthoryear{Madau \& Pozzetti}{2000}]{madau00} Madau P. \& Pozzetti L. 2000, MNRAS, 312, L9
\bibitem[\protect\citeauthoryear{Magorrian et al.}{1998}]{magorrian98} Magorrian J. et al. 1998, AJ, 115, 2285
\bibitem[\protect\citeauthoryear{Maraston}{2005}]{maraston05} Maraston C. 2005, MNRAS, 362, 799
\bibitem[\protect\citeauthoryear{Maraston et al.}{2006}]{maraston06} Maraston C., Daddi E., Renzini A., Cimatti A., Dickinson M., Papovich C., Pasquali A., Pirzkal N. 2006, ApJ, 652, 82
\bibitem[\protect\citeauthoryear{Marsden et al.}{2009}]{marsden09} Marsden G. et al. 2009, ApJ, 707, 1729
\bibitem[\protect\citeauthoryear{Matsumoto et al.}{2005}]{matsumoto05} Matsumoto T. et al., 2005, ApJ, 626, 31
\bibitem[\protect\citeauthoryear{Matsumoto et al.}{2015}]{matsumoto15} Matsumoto T., Kim M.~G., Pyo J., Tsumura K. 2015, ApJ, 807, 57
\bibitem[\protect\citeauthoryear{Matsuoka et al.}{2011}]{matsuoka11} Matsuoka Y., Ienaka N., Kawara K., Oyabu S. 2011, ApJ, 736, 119
\bibitem[\protect\citeauthoryear{Matsuura et al.}{2017}]{matsuura17} Matsuura S. et al. 2017, ApJ, 839, 7
\bibitem[\protect\citeauthoryear{Mattila et al.}{2017}]{mattila17} Mattila K., V\"{a}is\"{a}nen P., Lehtinen K., von Appen-Schnur G., Leinert C. 2017, MNRAS, in press
\bibitem[\protect\citeauthoryear{McAlpine et al.}{2013}]{mcalpine13} McAlpine K., Jarvis M.~J., Bonfield D.~G. 2013, A\&A, 544, A156
\bibitem[\protect\citeauthoryear{McDermid et al.}{2015}]{mcdermid15} McDermid R.~M. et al. 2015, MNRAS, 448, 3484
\bibitem[\protect\citeauthoryear{M\'{e}nard \& Fukugita}{2012}]{menard12} M\'{e}nard B. \& Fukugita M. 2012, ApJ, 754, 116
\bibitem[\protect\citeauthoryear{Merson et al.}{2016}]{merson16} Merson A. I., Baugh C. M., Gonzalez-Perez V., Abdalla F. B., Lagos C. d. P., Mei S. 2016, MNRAS, 456, 1681
\bibitem[\protect\citeauthoryear{Meurer et al.}{1999}]{meurer99} Meurer G.~R., Heckman T.~M., Calzetti D. 1999, ApJ, 521, 64
\bibitem[\protect\citeauthoryear{Mitchell et al.}{2014}]{mitchell14} Mitchell P. D., Lacey C. G., Cole S., Baugh C. M. 2014, MNRAS, 444, 2637
\bibitem[\protect\citeauthoryear{Moffett et al.}{2016a}]{moffett16} Moffett A.~J. et al. 2016a, MNRAS, 457, 1308
\bibitem[\protect\citeauthoryear{Moffett et al.}{2016b}]{moffett16b} Moffett A.~J. et al. 2016b, MNRAS, 462, 4336
\bibitem[\protect\citeauthoryear{Mullaney et al.}{2012a}]{mullaney12a} Mullaney J. R. et al. 2012a, MNRAS, 419, 9
\bibitem[\protect\citeauthoryear{Mullaney et al.}{2012b}]{mullaney12b} Mullaney J. R. et al. 2012b, ApJ, 753, L30
\bibitem[\protect\citeauthoryear{No\"{e}l et al.}{2013}]{noel13} No\"{e}l N.~E.~D., Greggio L., Renzini A., Carollo C.~M., Maraston C. 2013, ApJ, 772, 58
\bibitem[\protect\citeauthoryear{Padovani et al.}{2015}]{padovani15} Padovani P., Bonzini M., Kellermann K. I., Miller N., Mainieri V., Tozzi P. 2015, MNRAS, 452, 1263
\bibitem[\protect\citeauthoryear{Palanque-Delabrouille et al.}{2016}]{palanque16} Palanque-Delabrouille~N. et al. 2016, A\&A, 587, A41
\bibitem[\protect\citeauthoryear{Partridge \& Peebles}{1967a}]{partridge67a} Partridge R.~B. \& Peebles P.~J.~E. 1967, ApJ, 147, 868
\bibitem[\protect\citeauthoryear{Partridge \& Peebles}{1967b}]{partridge67b} Partridge R.~B. \& Peebles P.~J.~E. 1967, ApJ, 148, 377
\bibitem[\protect\citeauthoryear{Prugniel \& Soubiran}{2001}]{prugniel01} Prugniel P. \& Soubiran C. 2001, A\&A, 369, 1048
\bibitem[\protect\citeauthoryear{Puget et al.}{1996}]{puget96} Puget J.-L., Abergel A., Bernard J.-P., Boulanger F., Burton W.~B., Desert F.-X., Hartmann D. 1996, A\&A, 308, L5
\bibitem[\protect\citeauthoryear{Richards et al.}{2006}]{richards06} Richards G. et al., 2006, AJ, 131, 2766
\bibitem[\protect\citeauthoryear{Rieke et al.}{2009}]{rieke09} Rieke G.~H., Alonso-Herrero A., Weiner B.~J., P\'{e}rez-Gonz\'{a}lez P.~G., Blaylock M., Donley J.~L., Marcillac D. 2009, ApJ, 692, 556
\bibitem[\protect\citeauthoryear{Robotham et al.}{2011}]{robotham11} Robotham A.~S.~G. et al. 2011, MNRAS, 416, 2640
\bibitem[\protect\citeauthoryear{Robotham et al.}{2014}]{robotham14} Robotham A.~S.~G. et al. 2014, MNRAS, 444, 3986
\bibitem[\protect\citeauthoryear{Robotham et al.}{2017}]{robotham17} Robotham A.~S.~G., Taranu D.~S., Tobar R., Moffett A., Driver S.~P. 2017, MNRAS, 466, 1513
\bibitem[\protect\citeauthoryear{Rodighiero et al.}{2015}]{rodighiero15} Rodighiero G. et al. 2015, ApJ, 800, L10
\bibitem[\protect\citeauthoryear{Rosario et al.}{2012}]{rosario12} Rosario D.~J. et al. 2012, A\&A, 545, A45
\bibitem[\protect\citeauthoryear{Rowlands et al.}{2012}]{rowlands12} Rowlands K. et al. 2012, MNRAS, 419, 2545
\bibitem[\protect\citeauthoryear{Savage \& Mathis}{1979}]{savage79} Savage B.~D. \& Mathis J.~S. 1979, ARA\&A, 17, 73
\bibitem[\protect\citeauthoryear{Schaye et al.}{2015}]{schaye15} Schaye J. et al. 2015, MNRAS, 446, 521
\bibitem[\protect\citeauthoryear{Scoville et al.}{2007}]{scoville07} Scoville N. et al. 2007, ApJS, 172, 1
\bibitem[\protect\citeauthoryear{Siebenmorgen et al.}{2015}]{siebenmorgen15} Siebenmorgen R., Heymann F., Efstathiou A. 2015, A\&A, 583, A180
\bibitem[\protect\citeauthoryear{Silk \& Rees}{1998}]{silk98} Silk J. \& Rees M. J. 1998, A\&A, 331, L1
\bibitem[\protect\citeauthoryear{Smith et al.}{2012}]{smith12} Smith M.~W.~L. et al. 2012, ApJ, 748, 123
\bibitem[\protect\citeauthoryear{Smol\u{c}i\'{c} et al.}{2009}]{smolcic09} Smol\u{c}i\'{c} V., et al., 2009, ApJ, 696, 24
\bibitem[\protect\citeauthoryear{Somerville et al.}{2008}]{somerville08} Somerville R. S., Hopkins P. F., Cox T. J., Robertson B. E., Hernquist L. 2008, MNRAS, 391, 481
\bibitem[\protect\citeauthoryear{Somerville et al.}{2012}]{somerville12} Somerville R.~S., Gilmore R.~C., Primack J.~R., Dom\'{i}nguez A. 2012, MNRAS, 423, 1992
\bibitem[\protect\citeauthoryear{Springel et al.}{2005}]{springel05} Springel V. et al, 2005, Nature, 435, 629
\bibitem[\protect\citeauthoryear{Stanley et al.}{2015}]{stanley15} Stanley F., Harrison C.~M., Alexander D.~M., Swinbank A.~M., Aird J.~A, Del Moro A., Hickox R.~C., Mullaney J.~R. 2015, MNRAS, 453, 591
\bibitem[\protect\citeauthoryear{Symeonidis et al.}{2013}]{symeonidis13} Symeonidis M. et al. 2013, MNRAS, 431, 2317
\bibitem[\protect\citeauthoryear{Taylor et al.}{2011}]{taylor11} Taylor E.~N. et al. 2011, MNRAS, 418, 1587
\bibitem[\protect\citeauthoryear{Thilker et al.}{2007}]{thilker07} Thilker D.~A. et al. 2007, ApJS, 173, 538
\bibitem[\protect\citeauthoryear{Totani et al.}{2001}]{totani01} Totani T., Yoshii Y., Iwamuro F., Maihara T., Motohara K. 2001, ApJ, 550, L137
\bibitem[\protect\citeauthoryear{Tremonti et al.}{2004}]{tremonti04} Tremonti C.~A. et al., 2004, ApJ, 613, 898
\bibitem[\protect\citeauthoryear{Treister \& Urry}{2006}]{triester06} Triester E. \& Urry C.~M. 2006, ApJ, 652, L79
\bibitem[\protect\citeauthoryear{Ueda et al.}{2014}]{ueda14} Ueda Y., Akiyama M., Hasinger G., Miyaji T., Watson M.~G. 2014, ApJ, 786, 104
\bibitem[\protect\citeauthoryear{vanden Berk et al.}{2001}]{vandenberk01} vanden Berk D.~E. et al. 2001, AJ, 122, 549
\bibitem[\protect\citeauthoryear{Viero et al.}{2013}]{viero13} Viero M.~P. et al. 2013, ApJ, 779, 32
\bibitem[\protect\citeauthoryear{Viero et al.}{2015}]{viero15} Viero M.~P. et al. 2015, ApJ, 809, L22
\bibitem[\protect\citeauthoryear{Wilkins et al.}{2008}]{wilkins08} Wilkins S.~M., Trentham N., Hopkins A.~M. 2008, MNRAS, 385, 687
\bibitem[\protect\citeauthoryear{Wright et al.}{2017}]{wright17} Wright A.~H. et al. 2017, MNRAS, in press
\bibitem[\protect\citeauthoryear{Wright et al.}{2016a}]{wright16a} Wright A.~H. et al. 2016, MNRAS, 460, 765
\bibitem[\protect\citeauthoryear{Xu et al.}{2005}]{xu05} Xu C. et al. 2005, ApJ, 619, L11
\bibitem[\protect\citeauthoryear{Zahid et al.}{2011}]{zahid11} Zahid H. J., Kewley L. J., Bresolin F., 2011, ApJ, 730, 137
\bibitem[\protect\citeauthoryear{Zavala et al.}{2017}]{zavala17} Zavala J.~A. et al. 2017, MNRAS, 464, 3369
\bibitem[\protect\citeauthoryear{Zemcov et al.}{2017}]{zemcov17} Zemcov M., Immel P., Nguyen C., Cooray A., Lisse C.~M., Poppe A.~R. 2017, Nature Communications, 8, 15003
\bibitem[\protect\citeauthoryear{Zubko et al.}{2004}]{zubko04} Zubko V., Dwek E., Arendt R.~G. 2004, ApJS, 152, 211
\end{thebibliography}
\end{document}